\documentclass[twocolumn,amsmath,amssymb,amsfonts,floatfix,showpacs]{revtex4}

\usepackage{color}
\usepackage{graphicx}
\usepackage{dcolumn}
\usepackage{bm}
\usepackage{soul}
\usepackage{bbm}
\usepackage{ulem}
\definecolor{orange}{rgb}{1,0.5,0}
\definecolor{darkgreen}{rgb}{0.1,0.6,0.2}
\usepackage[english]{babel}
\usepackage{amsmath}
\usepackage{gnuplottex} 
\usepackage{graphicx}
\usepackage{braket}
\newcommand{\sgn}{{\rm sgn}}

\newcommand{\bea}{\begin{eqnarray}}
\newcommand{\eea}{\end{eqnarray}}
\newcommand{\bt}{\textbf}

\newcommand{\phd}{\phantom{\dag}}
\newcommand{\ph}{\phantom{.}}

\newcommand{\noi}{\noindent}
\newcommand{\no}{\nonumber}

\begin{document}
\title{Interplay of topological phases in magnetic adatom-chains on top of\\ a Rashba superconducting surface}
\author{Andreas Heimes}
\email{andreas.heimes@kit.edu}
\author{Daniel Mendler}
\author{Panagiotis Kotetes}
\affiliation{Institut f\"ur Theoretische Festk\"orperphysik and DFG-Center for Functional Nanostructures (CFN), Karlsruhe Institute of Technology, 76128 Karlsruhe, Germany}

\vskip 1cm

\begin{abstract}
We investigate the topological properties and the accessible Majorana fermion (MF) phases ari\-sing in a hybrid device consisting of a chain of magnetic adatoms placed on the
surface of a conventional superconductor with Rashba spin-orbit coupling (SOC). By identifying the favored classical magnetic ground state of the adatom chain, we extract the
corresponding phase diagram which exhibits an interplay of ferromagnetic (FM), antiferromagnetic (AFM) and spiral orders. We determine the parameter regime for which the FM or
AFM phases dominate over the spiral and additionally become stable against thermal and quantum fluctuations. For the topological analysis we focus on the FM and AFM cases and
employ a low-energy effective model relying on Shiba bound states. We find that for both magnetic patterns the hybrid system behaves as a topological superconductor which can
harbor one or even two MFs per edge, due to chiral symmetry. As we show, the two magnetic orderings lead to qualitatively and quantitatively distinct topological features
that are reflected in the spatial profile of the MF wavefunctions. Finally, we propose directions on how to experimentally access the diverse MF phases by varying the adatom
spacing, the SOC strength, or the magnetic moment of the adatoms in consideration.  
\end{abstract}

\pacs{74.78.-w, 74.45.+c, 75.75.-c}

\maketitle

Materials with Rashba spin-orbit coupling (SOC) have recently attracted renewed attention due to their pi\-vo\-tal role for realizing artificial topological superconductors
(TSCs) which harbor Majorana fermions (MFs) \cite{KaneHasan,ZhangQi,AliceaReview,BeenakkerReview,KotetesClassi}. Early proposals involved materials with SOC, such as
topological insulators \cite{FuKane2008}, non-centrosymmetric SCs \cite{Sato}, and Rashba semiconductors \cite{SauSemi,AliceaSemi,Sau,Oreg}, which stimulated significant
experimental progress. Remar\-kab\-ly, a number of promising but yet not fully conclusive MF-signatures have been already reported in semiconductor-based heterostructures
\cite{Mourik,Deng,Furdyna,Heiblum}. The unsettled witnes\-sing of MFs \cite{Lee,Finck,Marcus} constitutes a strong motivation for engineering and testing alternative hybrid
devices. For in\-stance, platforms based on magnetic adatoms which can be manipulated and probed via spin-polarized and spatially-resolved scanning tunneling microscopy (STM)
techniques, appear capable of unambiguously revealing the presence of MFs. 

This new perspective opened the door for new MF devices based on magnetic adatoms on the surface of conventional superconductors. One finds implementations with magnetic
adatoms where the ordering is random \cite{Choy}, spiral \cite{Kjaergaard2012,Martin,Yazdani,Nakosai2013,Simon,Klinovaja2013,Franz,Pientka,JianLi,Ojanen},
antiferromagnetic (AFM) with SOC induced by the combination of Zeeman fields and supercurrents \cite{Heimes}, and ferromagnetic (FM) on top of a superconducting
surface with Rashba SOC \cite{Brydon,Hui2014}. Accor\-ding to very recent experimental findings \cite{Yazdani_Science}, MFs seem to indeed emerge in magnetic adatom hybrid
devices, where the ordering of the chain appears to be ferromagnetic. This type of ordering can lead to MFs only if Rashba SOC is present, arising from the broken inversion
associated with the Pb superconducting substrate. In fact, this is a plausible scenario for Pb which owes already a non-negligible intrinsic SOC \cite{PbSOC}. Evenmore, it has
been shown that the Rashba SOC arising in Pb quantum well structures can be considerably large and tunable \cite{Hugo2008,Yaji2009,Stomsky2013,Bihlmayer2007}. The related SOC
strength can even reach a correspon\-ding momentum splitting of the order of $\delta k \sim 0.05\,k_F$, where $k_F$ is the Fermi-momentum ($\hbar=1$). 
  
In this work we focus on a platform directly related to the recent experiment of Ref.~\cite{Yazdani_Science}. Specifically we consider a single chain consisting of
classical magnetic adatoms deposited on top of the surface of a SC with Rashba SOC. We first infer the energe\-tically favored classical magnetic order of the chain, out of
the possible FM, AFM and spiral profiles. Secondly, we investigate the topological properties of the arising engineered TSCs, particularly focusing on the topological FM and
AFM chains.

\begin{figure}[t]
\begin{center}
\includegraphics[width=\columnwidth]{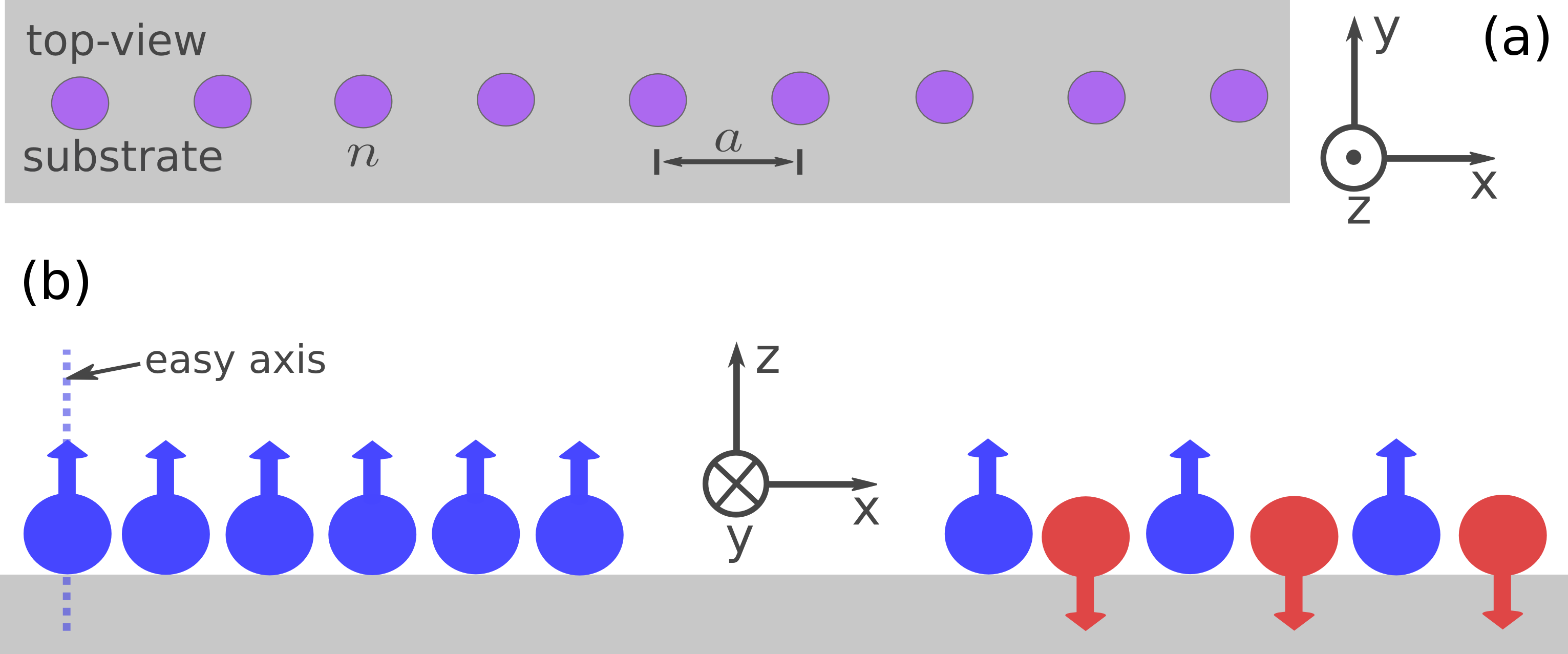}
\end{center}
\caption{
(a) Top view of a chain of adatoms placed on top of a superconducting surface with Rashba spin-orbit coupling (SOC). In the \textit{absence} of magnetism, the point
group symmetry of the hybrid structure, is $C_{2v}$, consisting of two reflection operations $\sigma_{xz}$ and $\sigma_{yz}$ (the index shows the mirror plane), and a $C_2$
rotation $(x,y,z)\rightarrow(-x,-y,z)$. (b) Side view of the hybrid structure. Crystal field effects (CFEs) violate spin rotational symmetry and favor an easy spin axis for
the magnetic ordering (here $z$ axis). On the other hand, SOC induces a Dzyaloshinsky-Moriya (DM) interaction. When the spin anisotropy dominates over the DM interaction, the
adatoms order in a ferromagnetic (FM) or antiferromagnetic (AFM) fashion, depending on the chain constant $a$. Otherwise, the spiral (SP) ordering prevails.
}
\label{fig::setup}
\end{figure}

In the first part of the manuscript, we explore the competition of the three aforementioned magnetic profiles by assuming identical adatoms owing a fixed spin $S$. The
magnetic atoms interact via a Ruderman-Kittel-Kasuya-Yosida (RKKY)-type superexchange \cite{RKKY}, which is mediated by the electrons of the SC. Due to the pre\-sen\-ce of
SOC, the resul\-ting superechange interaction is anisotropic and includes a Dzyaloshinsky-Moriya (DM) contribution \cite{DM}. The latter favors spiral orde\-ring which is stable against disorder if the SOC is sufficiently large \cite{Das Sarma}. On the other hand,
FM and AFM orders are stabilized by Ising-type anisotropy terms, induced by the crystal field of the substrate, which favor an easy axis for the magnetic ordering (see
Fig.~\ref{fig::setup}). By taking into account the various interactions, we extract the resul\-ting magnetic phase diagram by additionally varying the distance of the adatoms.
In this manner, our results address implementations with alternative substrates, either due to a different superconducting material or orientation of the surface involved.

In the second part, we focus on the topological pro\-per\-ties of these platforms, and concentrate on the FM and AFM cases. This is justified, as the findings of
Ref.~\cite{Yazdani_Science} indicate a strong Ising anisotropy, which as we show here, can additionally render the FM and AFM phases inert to quantum and thermal fluctuations
in spite of the one-dimensional character of the chain. For extracting the topological phase diagram, we first retrieve an effective low-energy model based on Shiba states
\cite{Shiba1968}, which constitute midgap electronic states of the SC localized at the sites of the adatoms. The symmetry properties of the system gives rise to a rich phase
diagram of MF-phases with 0, 1, or 2 MFs per chain edge. One can access the three phases via varying the adatom distance, the strength of the SOC and the value of the magnetic
moment. The phases with 2 MFs per chain edge are topologically protected by chiral \cite{Tewari and Sau,ChiralTanaka,SatoChiral,Hui2014,NOZeemanPK} symmetry, and they indeed
become accessible here for the parameters adopted. For illustrating the relevant mechanism driving the diverse topological phases, we identify the relevant gap closings in the
Shiba bandstructure, which provide insight for manipulating the MFs and tailoring the topological properties of these platforms. 

Our paper is structured as follows: In Sec.~\ref{section::magnetic_phases} we obtain the magnetic phase diagram for a magnetic chain on top of a metallic surface with Rashba
SOC. We consider that the magnetic adatoms interact via an RKKY interaction, while at the same time they experience a crystal field induced Ising anisotropy. In
Sec.~\ref{section::SOC_Superconductivity} we extend the previous analysis for the case of a superconducting substrate and discuss the modifications on the interplay of the
spiral, FM and AFM phases. In Sec.~\ref{section::topo_shiba}, we retrieve an effective one-dimensional low-energy Hamiltonian of the hybrid device for the FM and AFM
implementations, relying on Shiba bound states. In Secs.~\ref{section::FM} and \ref{section::AFM}, we extract the topological phase diagrams and study the arising MF
wavefunction characteristics, for the FM and AFM Shiba chains, respectively. Finally, we present our conclusions in Sec.~\ref{section::conclusion}.

\section{Magnetic phases of an adatom chain on a Rashba metallic surface}
\label{section::magnetic_phases}

In this section we discuss the favored ordering of a chain of magnetic atoms placed on top of a metallic surface with Rashba SOC. We first retrieve the RKKY superexchange
interaction between the magnetic atoms, which is mediated by the substrate electrons. By additionally taking into account an Ising-like anisotropy term due to the crystal
field, we retrieve the classical magnetic phase diagram, which consists of FM, AFM and spiral phases. Finally, we investigate the impact of quantum and thermal fluctuations
on the FM and AFM magnetic orders, and show that they are stable.

\subsection{RKKY interaction}
\label{section::SOC_2DEG}
 
We start with the Hamiltonian of a two dimensional metallic substrate with Rashba SOC: 
\bea
\label{eq::Rashba_Ham} H_{\rm metal}= \sum_{\bm k}  \psi_{\bm k}^\dagger\, h_{\bm k}\, \psi_{\bm k}\,,
\eea

\noindent where $h_{\bm k}=\xi_k + \alpha (\bm k \times \hat{\bm{z}})\cdot \bm \sigma$ is a $2\times 2$ matrix in spin-space and $\psi_{\bm k}^{\dag}=(c_{\bm
k\uparrow}^\dagger, \, c_{\bm k\downarrow}^\dagger)$ is the corresponding spinor. Furthermore, $c_{\bm k \sigma}^\dagger$ creates an electron with momentum $\bm k$ and spin
projection $\sigma$. The quadratic electronic dispersion, $\xi_{k}=k^2/2m-\mu$, can be linearized around the Fermi-momentum ($k_F=\sqrt{2m\mu}$), i.e. $\xi_{\bm k}
=v_F(k-k_F)$, where $k=|\bm k|$ and $v_F$ is the Fermi-velocity. The Hamiltonian $h_{\bm k}$ can be readily diagonalized via a
$\pi/2$-rotation about the $\hat{{\bm
k}}$-axis: 
\bea
e^{i\tfrac{\pi}{4} \hat{\bm k} \cdot \bm \sigma}h_{\bm k}\, e^{-i\tfrac{\pi}{4} \hat{\bm k} \cdot \bm \sigma}=\xi_k +\alpha k \sigma_z\,,
\eea

\noindent where $\hat{\bm k}=\bm k/k$. The respective eigenenergies are given by $\xi_{k \lambda} = \xi_k+\lambda \alpha k \approx v_F(k-k_\lambda)$, with $k_\lambda
\approx k_F(1-\alpha\lambda/v_F)$, corresponding to the two helicity bands $\lambda=\pm 1$. Thus, the effective momentum splitting $\delta k$ corresponds to a SOC strength
$\alpha=v_F\,\delta k/k_F$.

In order to proceed, we define the Matsubara Green's function in the helicity subspace: $g_\lambda(k,i\omega)=(i\omega - \xi_{k\lambda})^{-1}$ and with that we obtain
\bea
(i\omega-\xi_{k}-\alpha k\sigma_z)^{-1}=\sum_{\lambda=\pm}\frac{1+\lambda\sigma_z}{2}\ph g_\lambda(k,i\omega)\,.
\eea

\noindent According to the result above, the electronic Green's function is given by
\bea
G(\bm k,i\omega)&=&\sum_{\lambda=\pm}\frac{1+\lambda e^{-i\tfrac{\pi}{4} \hat{\bm k}\cdot \sigma}\sigma_z\,e^{i\tfrac{\pi}{4} \hat{\bm k}
\cdot \bm \sigma}}{2}\ph g_\lambda(k,i\omega)\no\\
&=&\sum_{\lambda=\pm}\frac{1+\lambda (\hat{\bm k}\times  \hat{\bm{z}})\cdot\bm \sigma}{2}\ph g_\lambda (k,i\omega)\,. \label{eq::GF_momentum}
\eea

At this point, we assume a certain arrangement for the magnetic adatoms on the metallic substrate. Here we consider classical spins $\bm S_i$ with magnitude $|\bm S_i|=S$,
placed at positions $\bm R_i=ia\hat{\bm{x}}$, with $i=1,\,...,\,N$. In addition, we consider that the interaction between the adatoms is driven by an exchange interaction
mediated by the conduction electrons of the substrate. The coupling between adatoms and conduction electrons can be parametrized by an exchange energy $J$, i.e.
\bea
\label{eq::spins_cond}
H_{J}=J\sum_{i=1}^N\iint\frac{d\bm k d\bm k'}{(2\pi)^2}\ph e^{-i(\bm k -\bm k ')\cdot \bm R_i} c_{\bm k\sigma}^\dagger
(\bm S_i\cdot \bm \sigma)_{\sigma\sigma'}c_{\bm k'\sigma'}\,.
\eea 

\noindent Given that $J$ is a small coupling constant and that the local modifications of the electronic spectrum in the substrate are negligible, we can follow a
standard one-loop expansion and obtain an effective spin-spin interaction. The so called RKKY interaction reads \cite{RKKY}
\begin{align}
\label{eq::RKKY}
H_{\rm RKKY}= -\frac{J^2}{2}\sum_{ij}\chi_{ij}^{\alpha\beta}S_i^\alpha S_j^\beta\,,
\end{align}

\noindent where the spin susceptibility can be derived using the Green's function given in Eq.~\eqref{eq::GF_momentum}:
\begin{align}
\label{eq::susceptibility}
\chi_{ij}^{\alpha \beta}=-{T}\sum_{\omega}{\rm Tr}_{\sigma}\left[\sigma^\alpha G(\bm R_{i}-\bm R_{j},i\omega) \sigma^\beta G(\bm R_{j}-\bm R_{i},i\omega)\right],
\end{align}

\noi where $G(\bm R,i\omega) = \int \frac{d \bm k}{(2\pi)^2}\, e^{i\bm k\cdot\bm R} G(\bm k,i\omega)$. In the follo\-wing we will consider a chain of adatoms with magnetic
moments placed along the $x$ direction. In Appendix~\ref{section::appendix_RKKY} we present in detail the steps which yield the well known result \cite{Imamura2004} for the
RKKY interaction: 
\bea
\label{eq::RKKY_2}
&&H_{\rm RKKY}=-m\left(\frac{Jk_F}{\pi}\right)^2\sum_{ij}\frac{\sin(2k_F|r_{ij}|)}{(2k_F r_{ij})^2}\no\\
&\times&\bigg\{\cos(2m\alpha r_{ij})\bm S_i\cdot\bm S_j+[1-\cos(2m\alpha r_{ij})]S_i^yS_j^y\qquad\no\\
&&\quad+\sin(2m\alpha r_{ij})\left(\bm S_i\times \bm S_j\right)_y\bigg\}\,,
\eea

\noindent where $\nu_F=m/2\pi{}$ is the density of states at the Fermi-level for each spin-band and $r_{ij}\equiv(i-j)a$. { Eq.~\eqref{eq::RKKY_2} holds in the
limit $k_Fa \gg 1$.} For vani\-shing SOC, we recover the usual spin rotationally invariant Heisenberg interaction, proportional to $\bm S_i\cdot \bm S_j$. On the other hand, a
finite SOC produces both an additional Ising interaction $S_i^yS_j^y$ and a DM interaction $(\bm S_i\times \bm S_j)_y$. Note, that the particular form for the RKKY interaction
could have been readily retrieved by considering all the bilinear spin-spin interaction terms, which are allowed by the $C_{2v}$ point group symmetry of the system in the
non-magnetic phase. 

We may write the RKKY interaction in a compact fashion, by taking into account that the rotation of a classical spin $\bm S_j$ by an angle $\theta_{ij}=2m\alpha r_{ij}$
with respect to $\bm S_i$, is given by 
\begin{align}
\bm S_j(\theta_{ij}) \equiv \cos(\theta_{ij})\bm S_j+\sin(\theta_{ij})\left(\hat {\bm y}\times\bm S_j\right)\,.
\end{align}

\noindent Thus we may rewrite Eq.~\eqref{eq::RKKY_2} as follows \cite{Imamura2004}:
\bea
H_{\rm RKKY}=-m\left(\frac{J k_F}{\pi}\right)^2\sum_{ij} \frac{\sin(2k_F|r_{ij}|)}{(2k_F r_{ij})^2}\ph\bm S_i\cdot\bm S_j(\theta_{ij}).\phd
\eea

\noindent This implies that the SOC would generally establish a spiral configuration, with a tilting angle $\theta_{i+1,i}=2ma\alpha$ between successive spins. However, the
substrate crystal field effects (CFEs), violate spin rotational invariance so that the magnetic moment of adatoms tends to point along the axis perpendicular to the surface
($z$ axis here). This anisotropy gives rise to an additional term appea\-ring in the total adatom Hamiltonian, which depends on the microscopic details of the substrate and
can generally assume a rather complicated form. However, here we will consider the simplest allowed term with the form
\bea
H_{\rm CFE}=-\frac{D}{2}\sum_i (S_i^z)^2\,,
\eea

\noi which accounts for the broken spin-rotational invariance. The parameter $D$ has been already estimated experimentally for some cases, by means of spin-polarized STM
\cite{Brune2009,Khajetoorians2012}.

\subsection{Classical magnetic ground state} \label{section::classical_ground_state}

\begin{figure}[t]
\begin{center}
\includegraphics[width=0.8\columnwidth]{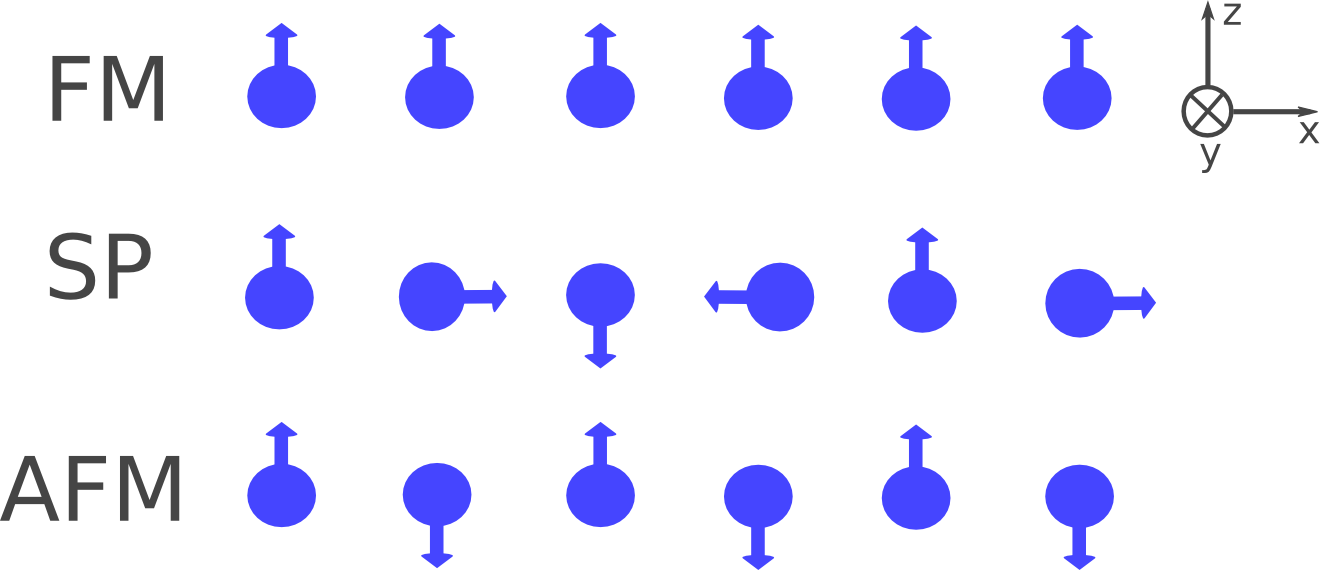}
\end{center}
\caption{Possible scenarios for the classical magnetic ground state: ferromagnetic (FM), spiral (SP) and antiferromagnetic (AFM) ordering.}
\label{fig::phases}
\end{figure}

In this paragraph, we discuss the competition of the possible magnetic phases of the adatom chain (see Fig.~\ref{fig::phases}), arising from the interplay of the SOC and the
CFEs. The former favors a spiral ordering while the latter, if large enough, can stabilize a FM or AFM ordering depending on the adatom spa\-cing. For the rest, we treat the
spins classically, thus assuming that $|\bm S_i|=S$ with a fixed magnitude $S$. In the classical limit: $S\rightarrow \infty$ whereas $J\rightarrow 0$, so that $JS$ remains
finite. Later we will discuss the stability of the classical ground state against quantum and thermal fluctuations. 

\begin{figure}[b]
\begin{center}
\includegraphics[width=\columnwidth]{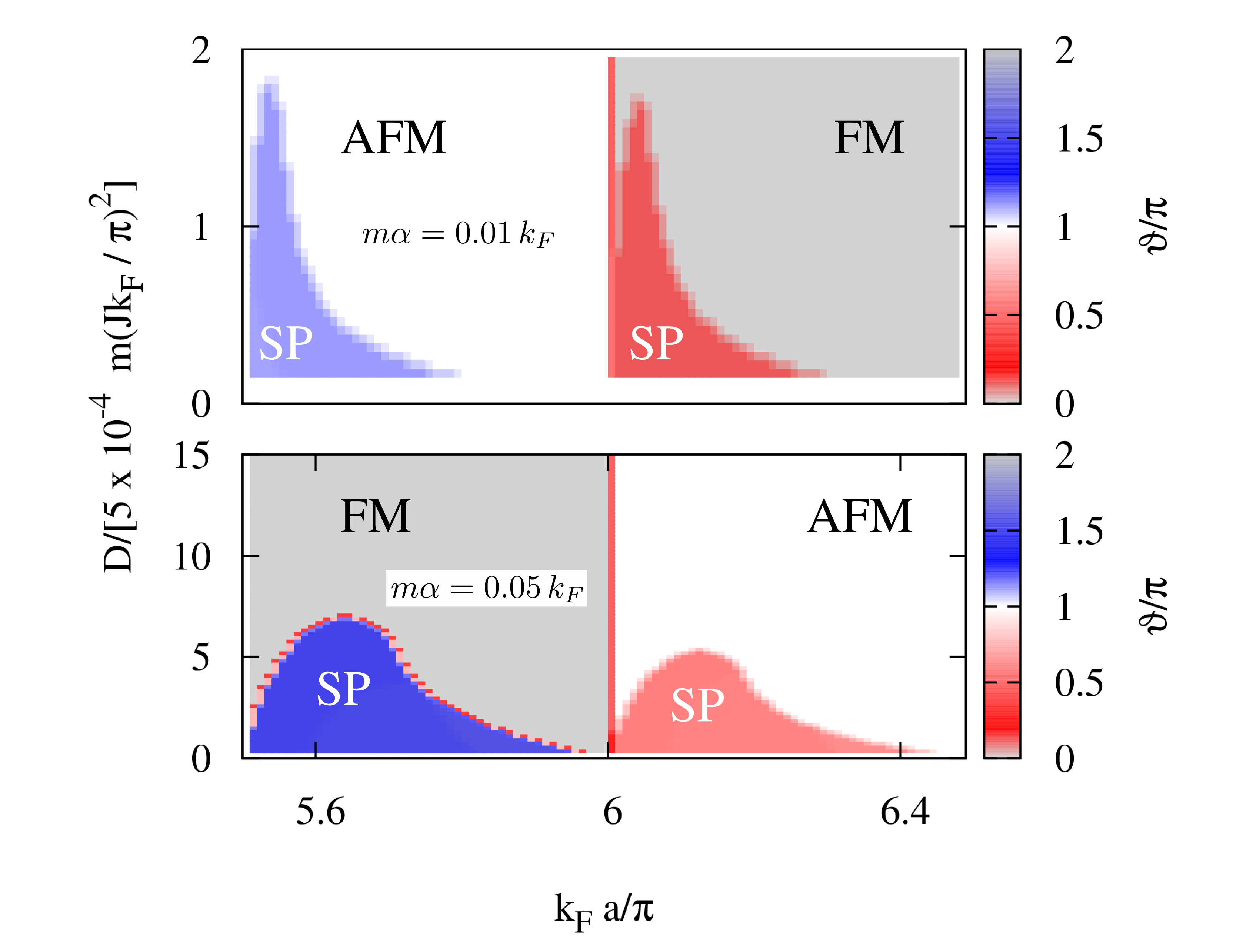}
\end{center}
\caption{Phase diagrams for the classical magnetic ground state. The parameter plane is defined by the adatom spacing $a$, and the rescaled strength $D$ of the crystal field
anisotropy. The presented diagrams were calculated for two values of the Rashba SOC strength $\alpha$. We find that large $\alpha$ coupling favors the spiral configuration,
whereas increasing the Ising anisotropy strength $D$ pins an easy axis ($z$) and promotes the FM and AFM phases.}
\label{fig::phase_1_fm_sp_afm}
\end{figure}

\begin{figure}[t]
\begin{center}
\includegraphics[width=\columnwidth]{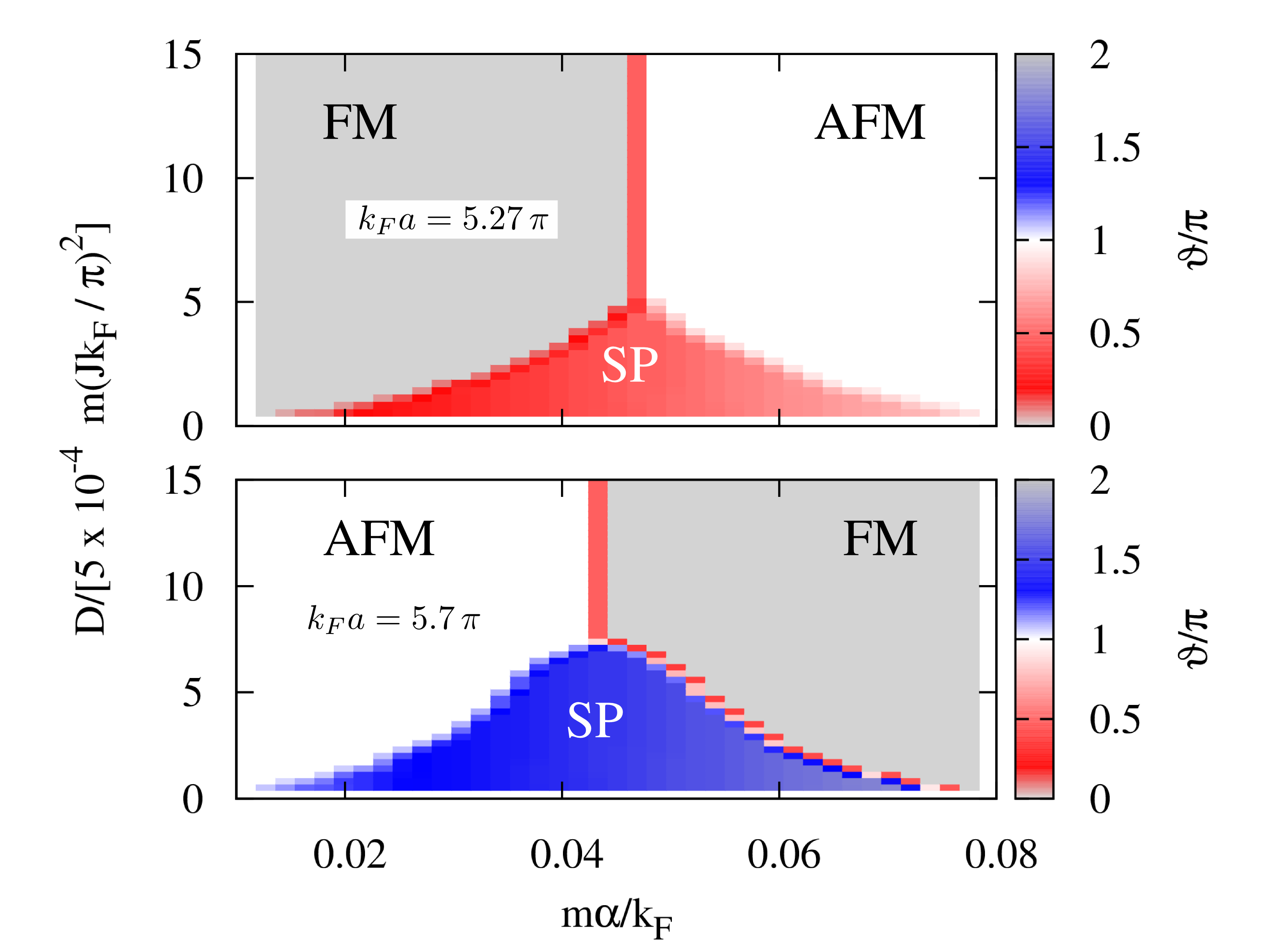}
\end{center}
\caption{Phase diagrams for the classical magnetic ground state. The parameter plane is defined by the rescaled SOC strength $\alpha$, and the rescaled strength $D$ of the
crystal field anisotropy. The presented diagrams were calculated for two values of the adatom spacing $a$. We observe stabilization of the FM or AFM phases for increasing CF
anisotropy. More importantly, tuning the SOC strength can \textit{tailor} the phase diagram leading to controllable switching between the FM and AFM phases.}
\label{fig::phase_2_fm_sp_afm}
\end{figure}

There are various ways to determine the classical ground state of the Hamiltonian $H_{\rm classical}=H_{\rm CFE}+H_{\rm RKKY}$. In this section we pursue a rather
qualitative discussion and we prefer to apply a trial configuration $\bm S_i(\vartheta)=S\sin(\vartheta i)\hat{\bm{x}}+S\cos(\vartheta i)\hat{\bm{z}}$, with the orientation
of the spins confined in the $xz$ plane. The latter form is fixed due to \bt{i.} the CFEs which energetically favor the appea\-rance of finite magnetization along the easy $z$
axis and \bt{ii.} the mixing of the $x$ and $z$ magnetization components induced by the DM interaction as an indirect result of the Rashba SOC. Therefore, the particular form
of the Hamiltonian implies that in the magnetic ground state, the spins are lying in the $xz$ plane. Under these conditions, the ground state is defined by the optimal value
of the angle $\vartheta$, which minimizes the classical Hamiltonian:
\bea
\label{eq::Hamiltonian_classical}
&&H_{\rm classical}(\vartheta)=-\frac{DS^2}{2}\sum_i \cos^2(\vartheta i)\\
&&-m\left(\frac{JSk_F}{\pi}\right)^2\sum_{ij}\frac{\sin(2k_F|r_{ij}|)}{(2k_F r_{ij})^2} \cos[({2m\alpha+\vartheta/a})r_{ij}]\,.\no
\eea

\noi We minimize this Hamiltonian with respect to $\vartheta$ for an infinite chain. In Fig.~\ref{fig::phase_1_fm_sp_afm} we see that depending on the relation between:
\bt{i}. exchange energy $JS$, \bt{ii.} CF anisotropy $D$, \bt{iii.} SOC strength $\alpha k_F$, and \bt{iv.} adatom spacing $a$, the classical ground state can assume a FM
($\vartheta=0$), AFM ($\vartheta=\pi$) or spiral configuration ($\vartheta\neq0,\pi$). The stronger the SOC, the stronger the CF anisotropy that has to be present, in order to
compensate the tendency of the system to form a spiral. Moreover, we observe that by tuning the SOC strength, as for instance by applying an electric field along the $z$ axis,
we can realize a FM $\leftrightarrow$ AFM quantum phase transition. This can be clearly seen in Fig.~\ref{fig::phase_2_fm_sp_afm} where the different phases are shown for two
different values of the atomic spacing. The particular characteristic of the phase diagram implies that electric fields can be also employed for altering the topological
properties of the Shiba chains yielding a rich landscape of MF phases. Finally, note that the arising phase transitions are first order.

\subsection{Thermal and quantum fluctuations}
  
In this paragraph we will investigate the robustness of the FM and AFM phases against thermal and quantum fluctuations. To this end we assume that the system resides either
in the FM or the AFM phase, with $\vartheta_\pm=0,\pi$ respectively. We will retrieve the dispersion of the magnetic fluctuations for each magnetic phase, by employing a
Holstein-Primakoff (HP) transformation \cite{HP}. In the limit of large $S$, the HP transformation reads  
\bea
S_j^x= (\pm 1)^j \sqrt{S/2}\ph(b_j^{\dag}+b_j)\,,\quad S_j^y=i\sqrt{S/2}\ph(b_j^{\dag}-b_j)\,,\no\\
S_j^z= (\pm 1)^j (S-b^\dagger_jb_j)\,,\qquad\qquad\qquad\phd
\eea

\noi with $b_i$ and $b_i^\dagger$ constituting bosonic operators which obey the commutation relation $[b_i,\,b_j^\dagger]=\delta_{ij}$, where the indices $i,j$ refer to the
sites of the adatoms. In addition, $\pm$ corresponds to the FM ($+$) and the AFM ($-$) cases. At this point, we effect this transformation on Eq.~\eqref{eq::RKKY_2} and
separate the resulting quantum Hamiltonian in orders $H^{(m)}_{\rm quantum}$ with respect to the operators $b_i$ and $b_i^\dagger$. The zeroth order of the quantum Hamiltonian
coincides with the classical ground state energy given by 
\begin{align} \label{eq::gs_energy}
H^{(0)}_{\rm quantum,\pm}\equiv H_{\rm classical,\pm}=-\frac{NDS^2}{2}-\sum_{ij} \Xi_{i-j}^{\pm,\alpha} S^2\,,\\
\Xi_{i-j}^{\pm,\alpha}\equiv (\pm1)^{i-j}m\left(\frac{Jk_F}{\pi}\right)^2\frac{\sin(2k_F|r_{ij}|)}{(2k_F r_{ij})^2}\cos(2m\alpha r_{ij})\,.
\end{align}

\noi The linear term $H^{(1)}_{\rm quantum}$ vanishes, whereas the bilinear term is given by 
\bea
&&H_{\rm quantum,\pm}^{(2)}= -\frac{S}{2}\sum_{ij} \big(\Xi_{i-j}^{\pm,\alpha}-\Xi_{i-j}^{+,0}\big) \big(b_i^\dagger b_j^\dagger +b_ib_j\big)\no \\
&&-\frac{S}{2}\sum_{ij} \big(\Xi_{i-j}^{\pm,\alpha}+\Xi_{i-j}^{+,0}\big) \big(b_i^\dagger b_j + b_j^\dagger b_i\big)\no \\
&&+S\sum_{ij} \Xi_{i-j}^{\pm,\alpha} \big(b_i^\dagger b_i + b_j^\dagger b_j \big)+ \frac{D}{2}(2S-1)\sum_i b_i^\dagger b_i\,.
\eea

\noi In momentum space the Hamiltonian reads
\bea
H_{\rm quantum,\pm}^{(2)}=\sum_q\left[\gamma_q^{(1)} \big(b_q^\dagger b_q + b_{-q}^\dagger b_{-q}\big)\right.\no\\
+\left.\gamma_q^{(2)}\big(b_q^\dagger b_{-q}^\dagger + b_{q} b_{-q}\big)\right]\,,
\eea

\noi with the combinations 
\begin{align*}
\gamma_q^{(1)}&=-\frac{S}{2}\left(\Xi_q^{\pm,\alpha} +\Xi_q^{+,0}\right)+\frac{D}{4}(2S-1) + {S}\Xi_{q=0}^{\pm,\alpha}\\
\gamma_q^{(2)} &= -\frac{S}{2}\left(\Xi_q^{\pm,\alpha} -\Xi_q^{+,0}\right) 
\end{align*}

\noi and $\Xi_q^{\alpha,\pm}=\sum_j\exp(iqj)\Xi_{j}^{\alpha,\pm}$. % and $\Xi_k^\pm = (\Xi_k^\alpha \pm \Xi_k^{\alpha=0})/2$. 
A bosonic Bogoliubov transformation $b_q = u_q \beta_q - v_q \beta_{-q}^\dagger$ with $u_q=\cosh\eta_q$, $v_q=\sinh\eta_q$ and $\tanh(2\eta_q)=
\gamma_q^{(2)}/\gamma_q^{(1)}$, immediately provides the eigenenergies of the spin wave modes,
\begin{align*}
\omega_q = \sqrt{\big[\gamma_q^{(1)}\big]^2 - \big[\gamma_q^{(2)}\big]^2}\,.
\end{align*}

\noi In order to investigate the stability of the FM and AFM phases, we calculate the sublattice magnetization, i.e.
\begin{align*}
M=-\frac{1}{N}\sum_{j=1}^N (\pm 1)^j\braket{S_j^z} = S-\frac{1}{N}\sum_q \braket{b_q^\dagger b_q}.
\end{align*}

\noi Using the Bogoliubov operators and by introducing the Bose-Einstein distribution $n_q=\braket{\beta_q^\dagger \beta_q}$, we obtain the deviation of the sublattice
magnetization from its ground-state value
\begin{align}
\label{eq::magnetization}
\Delta S=S- M=\frac{1}{N}\sum_q \left[ n_q u_q^2 + (1{+n_q})v_q^2\right]\,,
\end{align}

\begin{figure}[t]
\begin{center}
\includegraphics[width=\columnwidth]{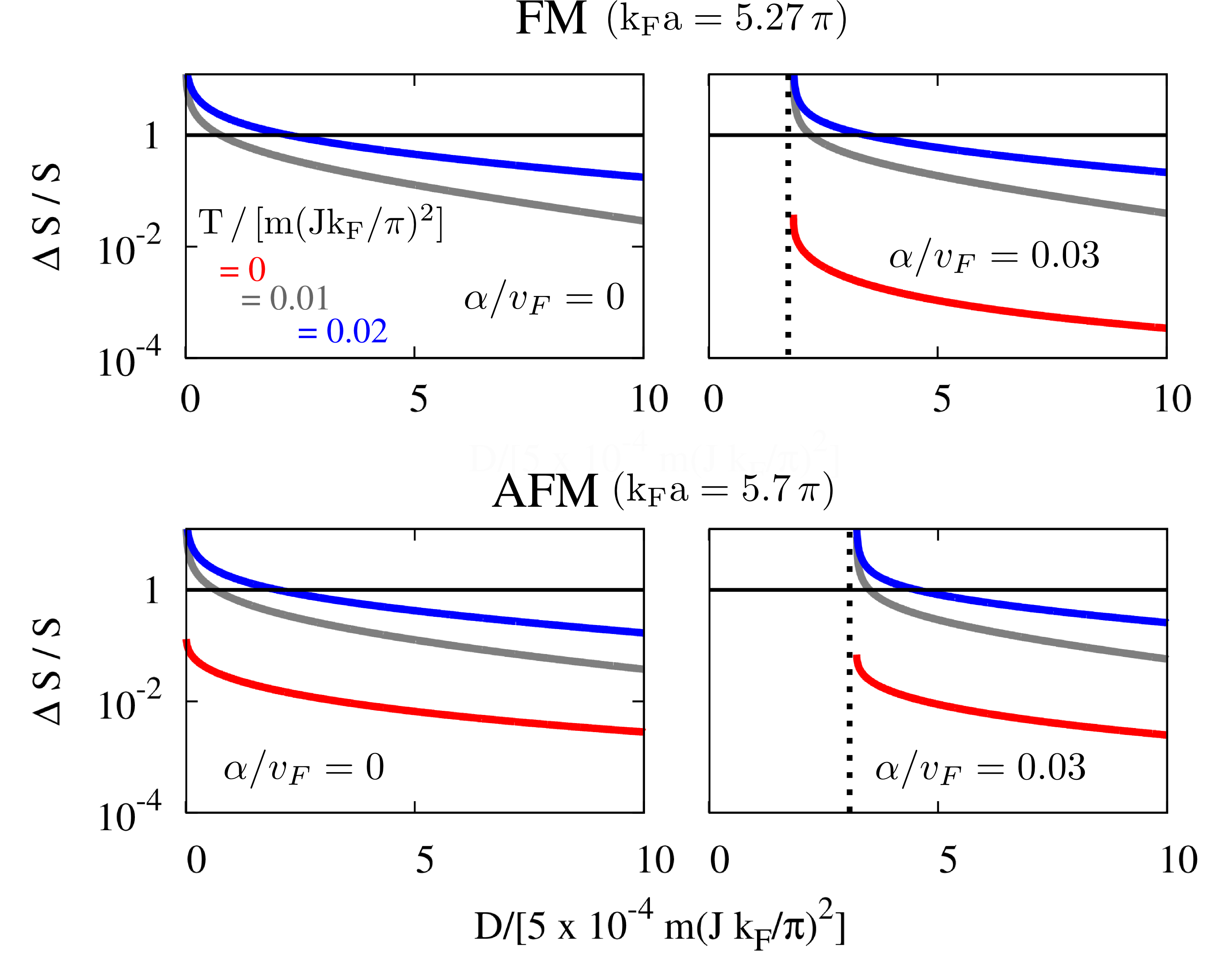}
\end{center}
\caption{Sublattice-magnetization measured from its ground-state value, $\Delta S$ ($S=15/2$), as a function of anisotropy $D$ for different tempe\-ra\-tures $T$ and SOC 
strength $\alpha$ for both the AFM and the FM configuration. Both phases become unstable for $D\rightarrow0$, since fluctuations become significant. The AFM phase always
exhibits both thermal and quantum fluctuations. In contrast, quantum fluctations appear in the FM case only when SOC is present. In both phases a sufficiently large, but
experimentally feasible, value for $D$ suppresses both types of fluctuations.}
\label{fig::holstein_primakoff}
\end{figure}

\noindent where we have assumed that $\sgn(\braket{S_1^z})=1$. In Fig.~\ref{fig::holstein_primakoff} we show $\Delta S$ for different
tem\-pe\-ra\-tures and SOC strength. For the atomic spin we use $S=15/2$ that has been realized in clusters of few magnetic atoms \cite{Khajetoorians2013}. We find that for
$T=0$ and $\alpha=0$ only the AFM configuration exhibits quantum fluctuations, which are absent in the FM case ($\Delta S=0$). In both cases we find that quantum as well as
thermal fluctuations are suppressed with increasing anisotropy $D$. For both AFM and FM configurations, the sum in Eq.~\eqref{eq::magnetization} diverges for finite
temperatures when taking the limit $D\rightarrow0$, and thus thermal fluctuations destroy the magnetic order. In STM experiments the nearest neighbor exchange energy as well
as the crystal field anisotropy can be measured. The next neighbor RKKY interaction of various metals is of the order \cite{Brune2009,Khajetoorians2012}
\begin{align}
\tilde J \equiv m\left(\frac{JSk_F}{\pi}\right)^2\frac{\sin(2k_Fa)}{(2k_F a)^2} \sim 0.1 \,\rm meV\,.
\end{align}

\noindent The crystal field anisotropy $D$ has been determined in Ref.~\cite{Khajetoorians2012} to be approximately given by $\sim 1 \,\rm meV$ or even larger
\cite{Brune2009}. In terms of the parameters $\tilde J$ and $D$ our calculation covers the parameter regime {$D/\tilde J \lesssim 10$}, which is consistent with the
aforementioned experimental realization. Furthermore, the so far explored temperatures are within the range $T\sim 0 -1\,$K, which are typical for the MF
experiments. As a conclusion, FM or AFM magnetic chains may be established, even in the presence of strong SOC without being destroyed by fluctuations.

\section{Magnetic phases of an adatom chain on a Rashba superconductor} \label{section::SOC_Superconductivity}
 
Here we extend the previous analysis in order to investigate the effect of superconductivity on the magnetic phase diagram. Once again, the magnetic adatoms interact via 
an RKKY interaction which is mediated by the electrons of the substrate superconductor, while they also feel a spin anisotropy due to the crystal field.

\subsection{Gor'kov-Nambu Green's function}

In this section we derive the Green's function for the superconducting substrate degrees of freedom in the pre\-sen\-ce of SOC. As before, we exclusively discuss Rashba SOC,
although other couplings between spin and momentum may be intrinsically present. This type of SOC can be engineered and can be considerably large for instance in quantum wells
\cite{Hugo2008,Yaji2009,Stomsky2013,Bihlmayer2007}. In fact, superconduc\-ting thin films of $\rm Pb$ feature both intrinsic and Rashba types of SOC. Starting from the
Hamiltonian of Eq.~\eqref{eq::Rashba_Ham} for a two-dimensional metallic surface with SOC, we consider here an additional spin singlet s-wave pairing term  $\Delta$ (here real
and positive)
\bea
H_{\rm sc}&=&\frac{1}{2} \sum_{\bm k} \psi_{\bm k}^\dagger\left[\xi_k \tau_z+\alpha \tau_z(\bm k \times \hat{\bm{z}})\cdot \tilde{\bm\sigma}-\Delta\tau_y\sigma_y\right]
\psi_{\bm k}\nonumber\\ 
&=&\frac{1}{2} \sum_{\bm k} \psi_{\bm k}^\dagger\,h_{\bm k}\psi_{\bm k}\,,  \label{eq::Ham_supercond}
\eea

\noi where the Pauli matrices $\bm{\tau}$ are defined in particle-hole space and $\psi_{\bm k}^\dagger = (c_{\bm k\uparrow}^\dagger,\,c_{\bm k\downarrow}^\dagger,\,c_{-\bm
k\uparrow},\,c_{-\bm k\downarrow})$ is the correspon\-ding spinor. Following the procedure of Sec.~\ref{section::SOC_2DEG} we perform a rotation, i.e. 
\begin{align*}
e^{i\tfrac{\pi}{4} \hat{\bm k} \cdot\tilde {\bm{\sigma}}} \,h_{\bm k} e^{-i\tfrac{\pi}{4} \hat{\bm k} \cdot \tilde{\bm{\sigma}}}=\xi_k \tau_z
+\alpha k \sigma_z-\Delta \tau_y\sigma_y\,.
\end{align*}

\noi Mind that the representation of the spin operator in the extended space is given by $\tilde{\bm{\sigma}}/2=(\tau_z\sigma_x,\sigma_y,\tau_z\sigma_z)/2$. By introducing
\bea
\tilde{g}_\pm(k,i\omega) = \big[(i\omega)^2-\Delta^2-\xi_{k\pm}^2\big]^{-1}\,,
\eea

\noi we obtain 
\begin{align*}
&\big[i\omega -  \xi_{k} \tau_z-\alpha k \sigma_z + \Delta \tau_y\sigma_y \big]^{-1}\\&=\sum_{\lambda=\pm}\frac{1+\lambda \tau_z\sigma_z}{2}(i\omega +
\xi_{k}\tau_z+\alpha k  \sigma_z- \Delta \tau_y\sigma_y)\tilde{g}_{\lambda}(k,i\omega)
\end{align*}

\noi and with the above, the electronic Gor'kov-Nambu Green's function becomes
\begin{align}
\hat G(\bm k,i\omega) &=e^{-i\tfrac{\pi}{4} \hat{\bm k} \cdot \tilde{\bm{\sigma}}}
\big[i\omega -  \xi_{k} \tau_z - \alpha k \sigma_z + \Delta \tau_y\sigma_y \big]^{-1} e^{i\tfrac{\pi}{4} \hat{\bm k} \cdot \tilde{\bm{\sigma}}} \nonumber\\
&=\sum_{\lambda=\pm} \frac{1+\lambda (\hat {\bm k}\times \hat {\bm z})\cdot\tilde{\bm{\sigma}}}{2}\frac{i\omega+\xi_{k\lambda}\tau_z}{(i\omega)^2-\Delta^2-\xi_{k\lambda}^2} 
\nonumber\\
&-\sum_{\lambda=\pm} \frac{1+\lambda (\hat {\bm k}\times \hat {\bm z})\cdot\tilde{\bm{\sigma}}}{2}\frac{\Delta \tau_y\sigma_y}{(i\omega)^2-\Delta^2-\xi_{k\lambda}^2}\,.
\label{eq::GNGF}
\end{align}

\noi Thus the presence of the Rashba SOC induces triplet pairing correlations \cite{Frigeri2004,patterns,GV}
\begin{align*}
\Delta(\hat {\bm k}\times \hat {\bm z})\cdot\tilde{\bm{\sigma}}\ph \tau_y\sigma_y = \Delta\left(\sin\varphi_{\bm{k}}\tau_x\sigma_z-\cos\varphi_{\bm{k}}\tau_y\right)\,,
\end{align*}

\noi where $\tan\varphi_{\bm{k}}=k_y/k_x$. The emergence of triplet correlations can be also understood within the theory of induced orders and patterns of coexisting phases
\cite{patterns,GV,CMR,Ce}. In this work, we assume only a local pairing interaction leading to a spin singlet superconducting order parameter $\Delta$ \cite{Tewari2011}, which is accompanied by
the triplet correlations above. However, in the presence of suitable non-local interactions which contribute to the above superconducting triplet channel, the s-wave singlet
and p-wave triplet order parameters \textit{necessarily} coexist at a microscopic level due to the SOC \cite{Frigeri2004,patterns,GV}. In the latter case, a p-wave spin
triplet order parameter has to be taken into account and determined self-consistently, as it can lead to modifications of the topological phase diagram \cite{KotetesClassi}.

In Eq.~\eqref{eq::GNGF} one can identify the electronic Gor'kov-Nambu Green's function
\begin{align*}
 G(\bm k,i\omega) &=\sum_{\lambda=\pm}\frac{1+\lambda (\hat {\bm k}\times \hat {\bm z})\cdot\bm{\sigma}}{2}\ph\frac{i\omega + \xi_{k\lambda}}{(i\omega)^2-\Delta^2-
 \xi_{k\lambda}^2}
\end{align*}

\noi and the anomalous one, 
\begin{align*}
F(\bm k,i\omega) &=\frac{\Delta}{2}\sum_{\lambda=\pm} \frac{i\sigma_y-\lambda(i\cos\varphi_{\bm{k}}+\sin\varphi_{\bm{k}}\sigma_z)}{(i\omega)^2-\Delta^2-\xi_{k\lambda}^2}\,.
\end{align*}

\noi By focusing on positions along the $x$ axis, i.e. $\bm{r}=r\hat{\bm{x}}$, we find
\begin{align}
\label{eq::sc_gf_2}
\hat G(r\hat{\bm x},i\omega)= \int \frac{d\bm k}{(2\pi)^2}\ph e^{ikr\cos\varphi_{\bm{k}}}\hat G({\bm k},i\omega)\qquad\qquad\qquad\qquad\ph\no\\
=\sum_{\lambda=\pm} \int_0^\infty \frac{d k\,k}{2\pi}\frac{J_0(kr)-i\lambda\sigma_y J_1(kr)}{2}\frac{i\omega+\tau_z \xi_{k \lambda} - \Delta \tau_y\sigma_y}
{(i\omega)^2 - \Delta^2 - \xi_{k\lambda}^2}.\no\\
\end{align}

\noi These expressions are valid if $\omega_D \gg v_F/r \sim E_F/k_Fr$, where $\omega_D$ is the Debye frequency. Mind that for $r=0$, we basically recover the electronic bulk
Green's function
\begin{align}
&\hat G(\bm{0},i\omega)= -\pi\nu_F \frac{i\omega - \Delta \tau_y\sigma_y}{\sqrt{\omega^2+\Delta^2}}\,.
\end{align}

\subsection{RKKY interaction}

In order to discuss the effective RKKY interaction mediated by the quasiparticles of a superconducting substrate with SOC, one can simply replace the Green's function
appearing in Eq.~\eqref{eq::susceptibility} by the one of Eq.~\eqref{eq::sc_gf_2} and the spin Pauli matrices $\sigma^\alpha$ and $\sigma^\beta$ by the corresponding
components of the spin Pauli vector in the new representation $\tilde{\bm{\sigma}}=(\tau_z\sigma_x,\sigma_y,\tau_z\sigma_z)$. 

For a superconducting substrate, the RKKY interaction owes an additional term, which does not arise in the case of metallic substrates. This distinct RKKY term is associated
with Shiba states \cite{Shiba1968}, appearing due to the presence of the magnetic adatoms on the superconduc\-ting surface. The latter constitute localized states at the
sites of the adatoms, with energies $\varepsilon_0$ which are smaller than the superconducting gap. In spite of the fact that the number of Shiba states is relatively small
compared to the number of the bulk accessible states, it has been recently shown that their contribution to the RKKY interaction can become important, favoring an AFM ordering
\cite{Yao2013}. Nonetheless, in order for the Shiba term to dominate over the bulk RKKY contribution the adatom spacing has to be rather long, since the former decays as
$(k_F r)^{-1}$ whereas the latter decays as $(k_F r)^{-2}$. The authors of Ref.~\cite{Yao2013} showed that the Shiba contribution dominates if the condition $k_Fr > \xi_0/r$
is fulfilled, holding for the material parameters and the atomic spacing ($r\sim 100 \rm nm$), which they focused on.
 
In stark contrast, here we assume an adatom spacing of the order of {1$\rm\, nm$} and a coherence length of $\xi_0 \sim 80\rm nm$. Therefore we find that $ k_Fr < \xi_0/r$ and
conclude that the Shiba bound state contribution is negligible in our case. Its inclusion would \textit{only} move the phase boun\-daries slightly deeper into the AFM region.
Moreover, since the main contribution to the RKKY interaction arises for ener\-gies quite above the gap $\Delta$, there is also no quantitative modification of the results
found pre\-viously in Sec.~\ref{section::magnetic_phases} for a normal metallic substrate. Therefore, the phase diagrams presented in Figs.~\ref{fig::phase_1_fm_sp_afm} and
\ref{fig::phase_2_fm_sp_afm}, also hold for the case of a superconducting substrate.

\section{Effective model for FM and AFM Shiba chains} \label{section::topo_shiba}

{As we explained in the previous paragraph, the contribution of the Shiba states to the RKKY interaction is unimportant in the case under consideration, and
thus their presence is irrelevant for deciding on the type of magnetic order which will develop in the adatom chain. However, the Shiba states are midgap states which
go\-vern the low-energy behavior of the electronic degrees of freedom. Therefore, in this section we proceed with investigating the characteristics of the Shiba states that
develop under the influence of background FM and AFM magnetic orderings of the adatom-spins $\bm{S}_i$, with $|\bm{S}_i|=S$, which have been stabilized by the effective RKKY
interaction originating solely from the continuum spectrum.} The magnetic exchange Hamiltonian now becomes
\bea
\label{eq::spins_cond_2}
H_{J}= \frac{1}{2}\sum_{j=1}^N\iint\frac{d\bm{k}d\bm{k}'}{(2\pi)^2}\ph e^{-i(\bm k -\bm k ')\cdot \bm R_j}\psi_{\bm k}^\dagger \,M_j\tau_z\sigma_z\,  \psi_{\bm k'}
\eea

\noi with $\bm R_i=ia\hat{\bm x}$. We introduced $M_j=JS(\pm 1)^j$, corresponding to FM ($+$) and AFM ($-$) ordering, respectively. In order to find the electronic spectrum 
we solve the Bogoliubov - de Gennes (BdG) equation \cite{Pientka,Yao2013,Heimes,Brydon,Balatsky2006,Flatte1997}
\bea
\label{eq::BdG_0}
\sum_j\int \frac{d\bm k'}{(2\pi)^2}\ph\hat G({\bm k},\varepsilon)e^{-i(\bm k -\bm k ')\cdot \bm R_j} M_j\tau_z\sigma_z\phi_{\bm k'}=\phi_{\bm k}\,,
\eea

\noi where the wavefunction $\phi_{\bm k}=(u_{\bm k\uparrow},u_{\bm k \downarrow},v_{\bm k\uparrow}, v_{\bm k\downarrow})^T$ contains the spin-dependent particle- and
hole-components $u$ and $v$. We assume that the continuum states are \textit{only slightly} affected by the presence of the magnetic atoms and thus assume the usual spectrum,
$E_{\bm k}=\sqrt{\Delta^2+\xi_{\bm k}^2}$, for the superconductor. By defining $\phi_j = (1/2\pi)\int d\bm k\, e^{i\bm k\cdot \bm R_j} \phi_{\bm k}$ we trace out the continuum
states and end up with the equation
\begin{align}
\label{eq::BdG_1}
\sum_j\bigg[V_i\delta_{ij} - V_i\hat G(r_{ij}\hat{\bm x},\varepsilon) V_j\bigg]\phi_j=0\,,
\end{align}

\noi where $V_j = M_j\tau_z \sigma_z$. By performing an expansion both in the normalized energy, $\varepsilon/\Delta$, as well as the couplings to higher order neighbors, 
we obtain from Eq.~\eqref{eq::sc_gf_2}

\bea
{\hat G}(\bm{0},\varepsilon)&\approx&-\pi \nu_F (\varepsilon/\Delta -  \tau_y\sigma_y)\,,\\
{\hat G}(r\hat{\bm x},\varepsilon)&\approx&{\cal G}^s(r)\tau_z+{\cal G}^a(r)\tau_z\sigma_y\no\\&+&{\cal F}^s(r)\tau_y\sigma_y+{\cal F}^a(r)\tau_y\,.
\eea

\noi The integrals in Eq.~\eqref{eq::sc_gf_2} together with the approximate form for the Bessel functions given in Eq.~\eqref{eq::bessel_approx}, yield the coefficients
\bea
\frac{{\cal G}^s(r)}{\pi\nu_F}&=&\cos(m\alpha r)\sin\big(k_F|r|-\tfrac{\pi}{4}\big){e^{-\tfrac{|r|}{\xi_0}}}\sqrt{\tfrac{2}{\pi k_F|r|}}\,,\phd\quad\\
\frac{{\cal F}^s(r)}{\pi\nu_F}&=&\cos(m\alpha r)\cos\big(k_F|r|-\tfrac{\pi}{4}\big){e^{-\tfrac{|r|}{\xi_0}}}\sqrt{\tfrac{2}{\pi k_F|r|}}\,,\phd\quad\\
\frac{{\cal G}^a(r)}{i\pi\nu_F}&=&\sin(m\alpha r)\sin\big(k_F|r|-\tfrac{\pi}{4}\big){e^{-\tfrac{|r|}{\xi_0}}}\sqrt{\tfrac{2}{\pi k_F|r|}}\,,\phd\quad\\
\frac{{\cal F}^a(r)}{i\pi\nu_F}&=&\sin(m\alpha r)\cos\big(k_F |r|-\tfrac{\pi}{4}\big){e^{-\tfrac{|r|}{\xi_0}}}\sqrt{\tfrac{2}{\pi k_F|r|}}\,,\phd\quad
\label{eq::GsGaFsFa}
\eea

\noi where $\xi_0$ is the coherence length of the superconductor. The indices $s$ and $a$ denote functions which are symmetric or anti-symmetric under inversion
$r\rightarrow-r$. With this we rewrite Eq.~\eqref{eq::BdG_1} in the form of a Schr\"odinger equation
\begin{align}
\label{eq::Schroedinger_eq}
\sum_j {\cal H}_{ij} \phi_j = {\varepsilon}\phi_i
\end{align}

\noi with the Hamiltonian
\bea
\label{eq::Hamiltonian_AFM}
{\cal H}_{ij}=\frac{\Delta}{\pi\nu_F M^2}\bigg[\left(\pi\nu_FM^2\tau_y\sigma_y-M_i\tau_z\sigma_z\right)\delta_{ij}\qquad\qquad\phd\no\\
+M_iM_j\left({\cal G}^s_{i-j}\tau_z-{\cal G}^a_{i-j}\tau_z\sigma_y+{\cal F}^s_{i-j}\tau_y\sigma_y-{\cal F}^a_{i-j}\tau_y\right)\bigg]\,,\quad
\eea

\noi where we have compactly denoted $f_{i-j}\equiv f(r_{ij})$ and set $M=JS$. The solution of Eq.~\eqref{eq::Schroedinger_eq} determines the energies and wavefunctions 
of the Shiba midgap states.

\section{Topological FM Shiba chain} \label{section::FM}

It has been well established, already from earlier proposals involving topological insulators \cite{FuKane2008} and semiconductors \cite{SauSemi,AliceaSemi,Sau,Oreg}, that the
combined presence of SOC, s-wave superconductivity and magnetism, can induce topological superconductivity. Note also that a recent symmetry classification
\cite{KotetesClassi} has presented further directions of how to combine these ingredients for engineering TSCs. Both cases of topological FM and AFM Shiba chains fall into
this classification scheme. In fact, the effective model for a topological Shiba chain (see also \cite{Brydon}) resembles previous continuum models describing TSCs using
nanowires with Rashba SOC \cite{Sau,Oreg}. In the present case, the effective Zeeman field is provided by the magnetic adatoms (classical here) and the SOC occurs due to the
involvement of a superconducting surface. However, there are also important differences. First, the perpendicular local magnetic field felt by the Shiba states is generally
less harmful for superconduc\-ti\-vi\-ty compared to a perpendicular magnetic field, due to the additional contribution of the orbital effects in the latter situation. In the
case of Shiba states, in spite of the fact that the superconducting gap becomes locally suppressed, it ge\-ne\-rally survives even when the magnetic exchange energy becomes
comparable to it \cite{Flatte1997}. In addition, note that the FM ordering is more likely to suppress superconductivity locally compared to the AFM ordering.

Moreover, another distinctive feature for the effective model of topological Shiba chains is that they incorporate triplet pairing correlations, which can ge\-ne\-ral\-ly
lead to a significant quantitative modification of the phase diagram (see Ref.~\cite{KotetesClassi}). In addition, the topological Shiba chain models are lattice
models involving higher order neighbor couplings, thus strongly depending on the adatom spacing. Consequently, one can not always restrict to a nearest neighbor model but
instead, depending on the ratio $\xi_0/a$, a large number of neighbors can become relevant. Evenmore, the inherent presence of additional chiral symmetries, leads to a rich
variety of topologically phases even with 2 MFs per edge. 

In the following paragraph, we first discuss the symmetries of a topological FM Shiba chain which are crucial for performing a topological classification of the accessible
MF phases. Furthermore, we extract the topological phase diagram and study numerically the MF wavefunctions for a finite chain, in order to discuss aspects related to the
experimental realization of this scenario.

\subsection{Symmetry classification}

As already mentioned, in the \textit{absence} of magnetism, the point group symmetry of the hybrid structure (as in Fig.~\ref{fig::setup}), consisting of the chain on top of
an infinite substrate surface, is $C_{2v}$. Since the effective model descri\-bing the Shiba chains is embedded in the two-dimensional geometry, rather than being a stictly
one dimensional system, it inherits the same point group properties. This is reflected in the $C_{2v}$ point group symmetry of the non-magnetic part of the Hamiltonian in
Eq.~\eqref{eq::Hamiltonian_AFM}. This point group consists of the: 
\begin{enumerate}
\item identity element $E:(x,y,z)\mapsto(x,y,z)$\,,
\item reflection operation $\sigma_{yz}:(x,y,z)\mapsto(-x,y,z)$\,,
\item reflection operation $\sigma_{xz}:(x,y,z)\mapsto(x,-y,z)$\,,
\item $z$ axis $\pi$-rotation $C_2:(x,y,z)\mapsto(-x,-y,z)$.
\end{enumerate}

\noi Note that for the effective Shiba state model of Eq.~\eqref{eq::Hamiltonian_AFM} only the inversion operation ${\cal I}:x\mapsto-x$ is accessible, and corresponds to
${\cal I}i=-i$ and ${\cal I}j=-j$, with $i,j$ denoting adatom sites. Therefore, within our spinor formalism the aforementioned symmetries are generated by the unitary
ope\-ra\-tors: $\hat{E}=I$, $\hat{\sigma}_{yz}=i\tau_z\sigma_x{\cal I}$, $\hat{\sigma}_{xz}=i\sigma_y$ and $\hat{C}_2=i\tau_z\sigma_z{\cal I}$. The term associated with the
presence of FM or\-de\-ring, $M_j\tau_z\sigma_z=M\tau_z\sigma_z$, transforms under the $C_{2v}$ elements in the following manner:
$\hat{\sigma}^{\dag}_{yz}\tau_z\sigma_z\hat{\sigma}_{yz}=-\tau_z\sigma_z$, $\hat{\sigma}^{\dag}_{xz}\tau_z\sigma_z\hat{\sigma}_{xz}=-\tau_z\sigma_z$,
$\hat{C}_2^{\dag}\tau_z\sigma_z\hat{C}_2=\tau_z\sigma_z$. Moreover, the FM chain is invariant under the action of the discrete translation operator, $\hat{t}_a$, which leads
to shift $i\mapsto i+1$, i.e. equal to the adatom spacing $a$. 

In contrast, the usual time-reversal ope\-ra\-tion ${\cal T}$ with generator $\hat{{\cal T}}=i\sigma_y\hat{{\cal K}}$, is broken as the FM term satisfies $\hat{{\cal
T}}^{\dag}\tau_z\sigma_z\hat{{\cal T}}=-\tau_z\sigma_z$. Here $\hat{{\cal K}}$ denotes the anti-unitary complex-conjugation ope\-ra\-tor. As it becomes evident from the above
relations, the FM term is invariant under the action of the following combined symmetry ope\-rations: ${\cal T}\sigma_{xz}$ and ${\cal T}\sigma_{yz}$, i.e. consisting of
operations under which the rest of the Hamiltonian is invariant. Usual\-ly, this type of symmetries are called \textit{hidden} symmetries \cite{SatoHidden,KotetesClassi}, as
they are a combination of symmetry operations which, separately, do not leave the Hamiltonian invariant. In the particular case only the action of the operator
$\hat{\Theta}\equiv\hat{\sigma}_{xz}\hat{{\cal T}}=\hat{{\cal K}}$, which coincides with the complex conjugation, leaves the total BdG Hamiltonian inva\-riant.

{Similarly to the usual time-reversal symmetry ope\-ra\-tor $\hat {\cal T}$, $\hat \Theta$ is also anti-unitary. We may thus call it a {\it ge\-ne\-ra\-li\-zed}
time-reversal symmetry ope\-ra\-tor \cite{KotetesClassi}. However, the operators differ in periodicity, i.e. $\hat{\cal T}^2=-I$ and $\hat{\Theta}^2=I$. The latter implies
that ${\cal T}$-symmetry will lead to a Kramers degeneracy, while $\Theta$-symmetry imposes a rea\-lity condition on the Hamiltonian without any Kramers pairs
\cite{Altland,KitaevClassi,Ryu}. In the case under consideration, the pre\-sen\-ce of $\Theta$-symmetry together with the built-in charge-conjugation symmetry of the BdG
Hamiltonian, effected by the operator $\hat{\Xi}\equiv\tau_x\hat{{\cal K}}$, give rise to the chiral symmetry operator $\hat{\Pi}\equiv\tau_x$. Thus although the usual
time-reversal symmetry is broken in our system, the presence of the aforementioned set of symmetries implies that the system resides in the BDI symmetry class, which in one
dimension can support topologically non-trivial phases characterized by a $\mathbb{Z}$ invariant \cite{Altland,KitaevClassi,Ryu}. The latter allows an integer number of MFs
per chain edge (see also \cite{Tewari and Sau,ChiralTanaka,SatoChiral,NOZeemanPK}). As we show in the next paragraph these topological phases are indeed accessible with the
particular system.}

\subsection{FM Shiba chain Hamiltonian}

In order to study the topological properties of a FM Shiba chain, we will transfer to momentum space, defined in the FM Brillouin zone (BZ) $k\in(-\pi/a,\pi/a]$. At this
point we introduce the corresponding BdG momentum space Hamiltonian, ${\cal H}_k={\cal H}_k^0+{\cal H}_k^{\rm m}$, consisting of the \bt{i.} non-magnetic ${\cal H}_k^0$ and
\bt{ii.} magnetic ${\cal H}_k^{\rm m}$ parts:
\bea\label{eq::HFMk}
{\cal H}_k^0&=&t_k\tau_z-v_k\tau_z\sigma_y+(\Delta+{\cal D}_k)\tau_y\sigma_y-d_k\tau_y\,,\quad\\
{\cal H}_k^{\rm m}&=&-{\cal B}\tau_z\sigma_z\,,
\eea

\noi where we have introduced ${\cal B}=\Delta/(\pi\nu_F JS)$ and
\bea
t_k&=&\sum_{\delta=1}^{\infty}t_{\delta}\cos(\delta ka)\quad{\rm with}\quad t_{\delta}=\frac{2\Delta}{\pi\nu_F}\ph{\cal G}^s_{\delta}\,,\\
v_k&=&\sum_{\delta=1}^{\infty}v_{\delta}\sin(\delta ka)\quad{\rm with}\quad v_{\delta}=\frac{2\Delta}{i\pi\nu_F}\ph{\cal G}^a_{\delta}\,,\\
{\cal D}_k&=&\sum_{\delta=1}^{\infty}{\cal D}_{\delta}\cos(\delta ka)\quad{\rm with}\quad{\cal D}_{\delta}=\frac{2\Delta}{\pi\nu_F}\ph{\cal F}^s_{\delta}\,,\\
d_k&=&\sum_{\delta=1}^{\infty}d_{\delta}\sin(\delta ka)\quad{\rm with}\quad d_{\delta}=\frac{2\Delta}{i\pi\nu_F}\ph{\cal F}^a_{\delta}\,.\qquad
\eea

\noi The above Hamiltonian acts on the wavefunction $\phi_{k}=(u_{k\uparrow},u_{k \downarrow},v_{k\uparrow}, v_{k\downarrow})^T$. In addition, $t_{\delta}$ corresponds to
the $\delta$-order nearest neighbor hopping, $v_{\delta}$ corresponds to the $\delta$-order nearest neighbor SOC, ${\cal D}_{\delta}$ to the $\delta$-order nearest neighbor
extended s-wave spin-singlet superconducting gap and $d_{\delta}$ to the $\delta$-order nearest neighbor spin-triplet superconducting gap oriented along the $y$ axis. 

\subsection{Topological invariant}

For exploring the topological phase diagram, we reside on the presence of chiral symmetry $\hat{\Pi}=\tau_x$ and block off-diagonalize the BdG Hamiltonian \cite{Tewari and
Sau,Ryu}, via a rotation about the $\tau_y$ axis effected by the unitary transformation $(\tau_z+\tau_x)/\sqrt{2}$. We obtain  
\begin{align}
{\cal H}_k'=
\begin{pmatrix}0 & A_k\\A^{\dag}_k & 0\end{pmatrix}\,.
\end{align}

\noi The upper block off-diagonal block is given by
\bea
A_k=t_k-id_k-{\cal B}\sigma_z-[v_k-i(\Delta+{\cal D}_k)]\sigma_y\,.
\eea

\noi The determinant of $A_k$ is a complex number and reads 
\bea
{\rm Det}[A_k]&=&t_k^2+(\Delta+{\cal D}_k)^2-{\cal B}^2-d_k^2-v_k^2\no\\&+&2i[v_k(\Delta+{\cal D}_k)-t_kd_k]\,.
\eea

\noi With the vectors $\bm{g}_k=({\rm Re}~{\rm Det}[A_k],{\rm Im}~{\rm Det}[A_k],0)$ and $\hat{\bm{g}}_k=\bm{g}_k/|\bm{g}_k|$, the related $\mathbb{Z}$ topological invariant
is defined by the winding number \cite{Volovik,Heimes}
\bea \label{eq::winding_1}
{\cal N}=\frac{1}{2\pi}\int_{BZ} dk \, \left(\hat{\bm{g}}_k\times \frac{\partial\hat{\bm{g}}_k}{\partial k}\right)_z\,.
\eea

\noi {Note that by considering a $\mathbb{Z}$ classification which also takes into account phases with 2 MFs, we manage to go beyond the study of a FM Shiba chain
performed in Ref.~\cite{Brydon}, which assumed a $\mathbb{Z}_2$ classification and thus restricted to the cases with 1 MF per edge.}

\subsection{Topological phase diagram -- Results}

\begin{figure}[b]
\begin{center}
\includegraphics[width=\columnwidth]{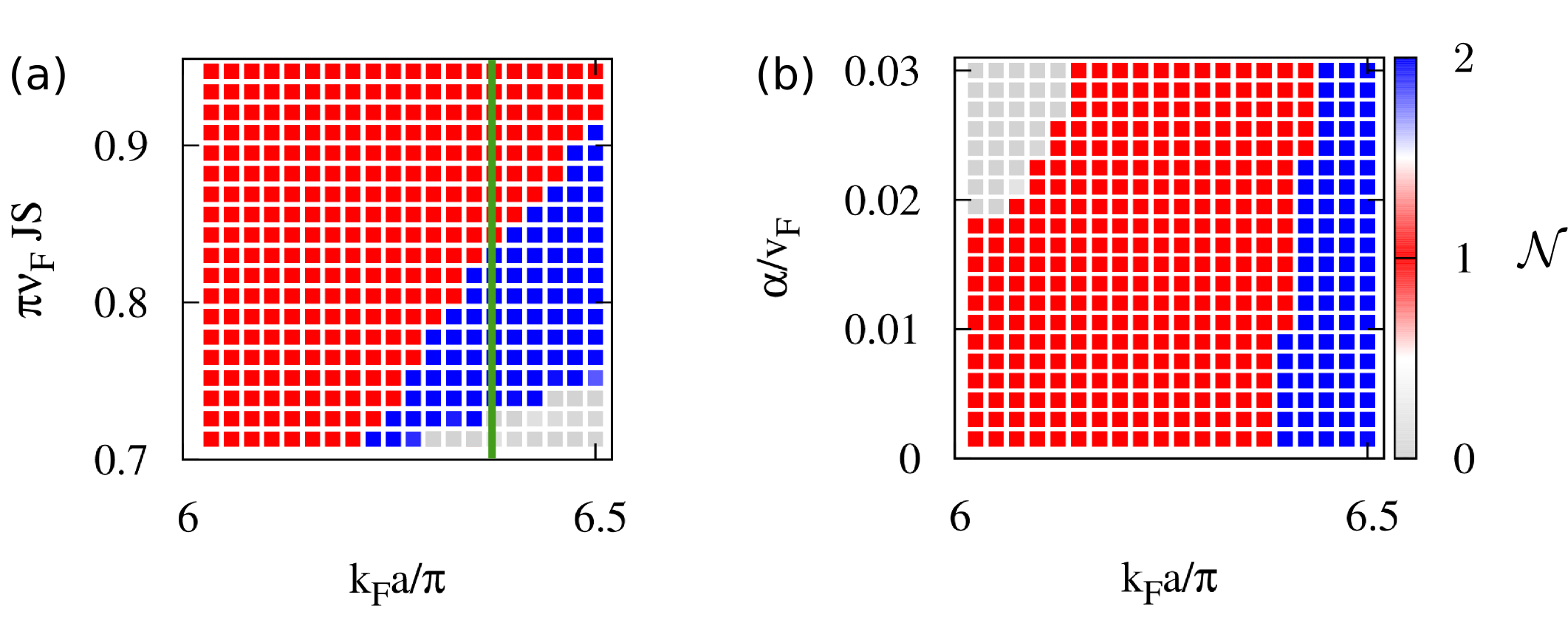}
\end{center}
\caption{The $\mathbb{Z}$ topological invariant (winding number) as defined in Eq.~\eqref{eq::winding_1}, for varying adatom spacing $a$ and (a) magnetic exchange energy $JS$ 
($\alpha=0.01\,v_F$) or (b) normalized SOC strength $\alpha$ ($\pi\nu_FJS=0.85$). In both cases, we find topological phases harboring 1 or 2 MFs per chain edge. 
Note in (a) that tuning the magnetic exchange energy can be used to switch between 1 and 2 MF phases. When close to the phase boundary, the latter could be for instance
achieved by applying a weak perpendicular Zeeman field. In (b) we observe that for an infinitessimally small SOC strength, both 1 and 2 MF phases are accessible. This is
anticipated for the single MF phase where $\alpha$ does not enter the topological criterion, but quite remarkably, it also takes place for the 2 MF situation. Here electrical
tuning of $\alpha$ can be used for realizing topological quantum phase transitions.
}
\label{fig::phase_fm}
\end{figure}

In Fig.~\ref{fig::phase_fm} we show the winding number as a function of the adatom spacing $a$, magnetic exchange energy $JS$ and SOC strength $\alpha$. Phases with zero, one
or two MFs per edge are accessible. When the ground state of the system resides in a phase near a boundary of the topological phase diagram, one can employ a weak
perpendicular Zeeman (electric) field to tune the magnetic exchange energy (SOC strength) in order to achieve transitions between phases with different number of MFs. We
additionally observe in Fig.~\ref{fig::phase_fm}(b) that the phase diagram exhibits MF bound states even for \textit{very small} values of $\alpha$. This is similar to the
nanowire case \cite{Sau,Oreg}, where $\alpha$ mainly determines the spatial profile and localization of the MFs at the edges of a finite system. Interestingly, this also 
holds for the case of 2 MFs per edge. 

\begin{figure}[t]
\begin{center}
\includegraphics[width=\columnwidth]{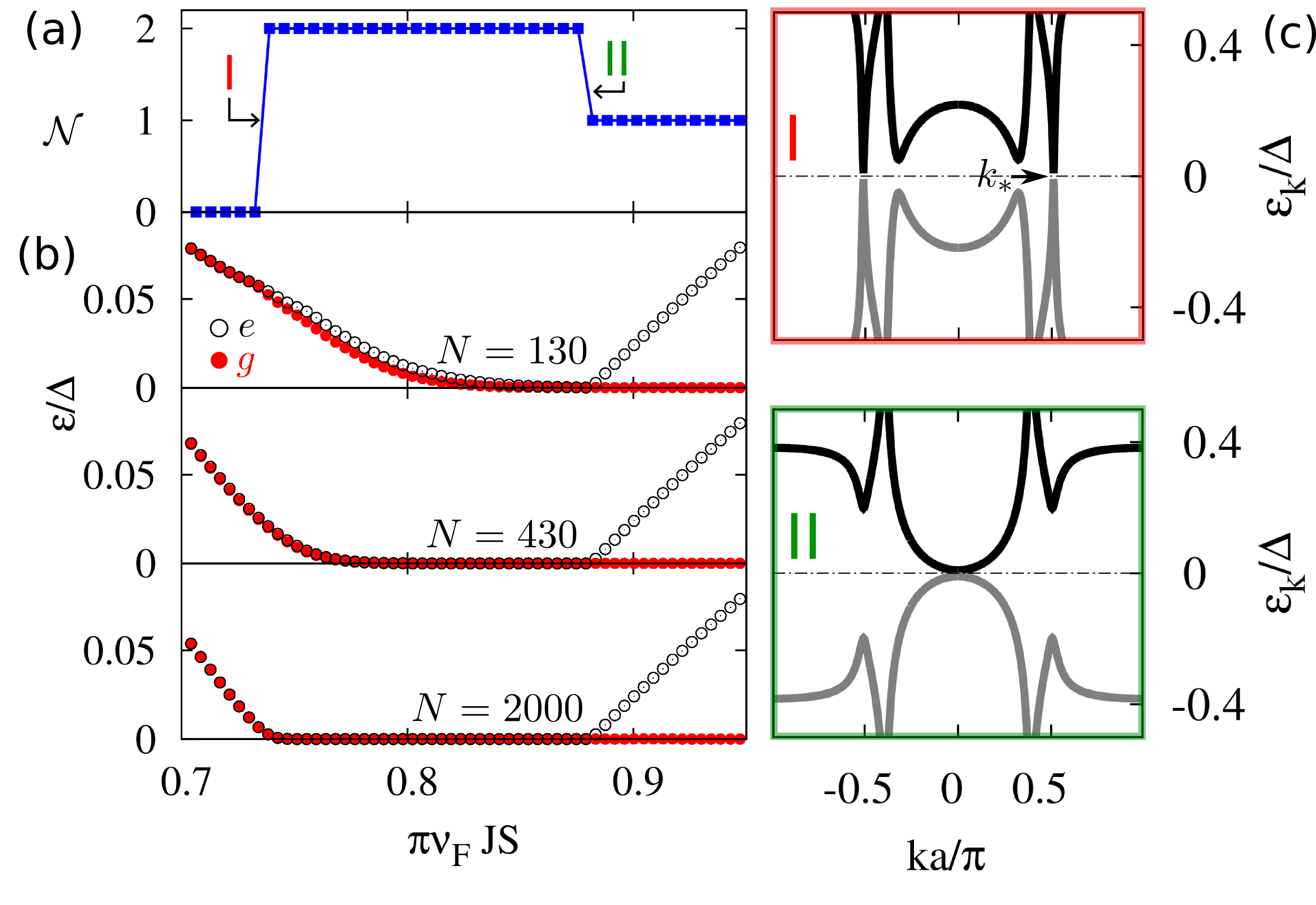}
\end{center}
\caption{(a) The $\mathbb{Z}$ topological invariant (winding number) as defined in Eq.~\eqref{eq::winding_1} (blue) depending on the magnetic exchange energy $JS$ along the
green line in Fig.~\ref{fig::phase_fm}(a). (b) The two lowest positive eigenenergies in red and black for three different lengths of the chain ($N=130$, $N=430$ and $N=2000$).
Note that in order to obtain well localized MF bound states and validate the bulk-boundary correspondence predictions, quite long chains are required. This is particularly the
case for the transition to the 2 MF phase, which is protected by chiral symmetry. In panel (c) we show the corresponding gap closings of the energy dispersions
$\varepsilon_k$, occuring exactly at the phase transition points {\color{red} I} and {\color{darkgreen} II}. The transition ({\color{red} I}) ${\cal N}=0\rightarrow2$ arises
from gap closings at the non-inversion-symmetric points $\pm k_*$, connected to each other by inversion. Instead, the transition ({\color{darkgreen} II}) ${\cal
N}=2\rightarrow1$ arises due to a gap closing at the inversion symmetric momentum $k=0$.
}  
\label{fig::fm_finite}
\end{figure}

\begin{figure}[t]
\begin{center}
\includegraphics[width=0.85\columnwidth]{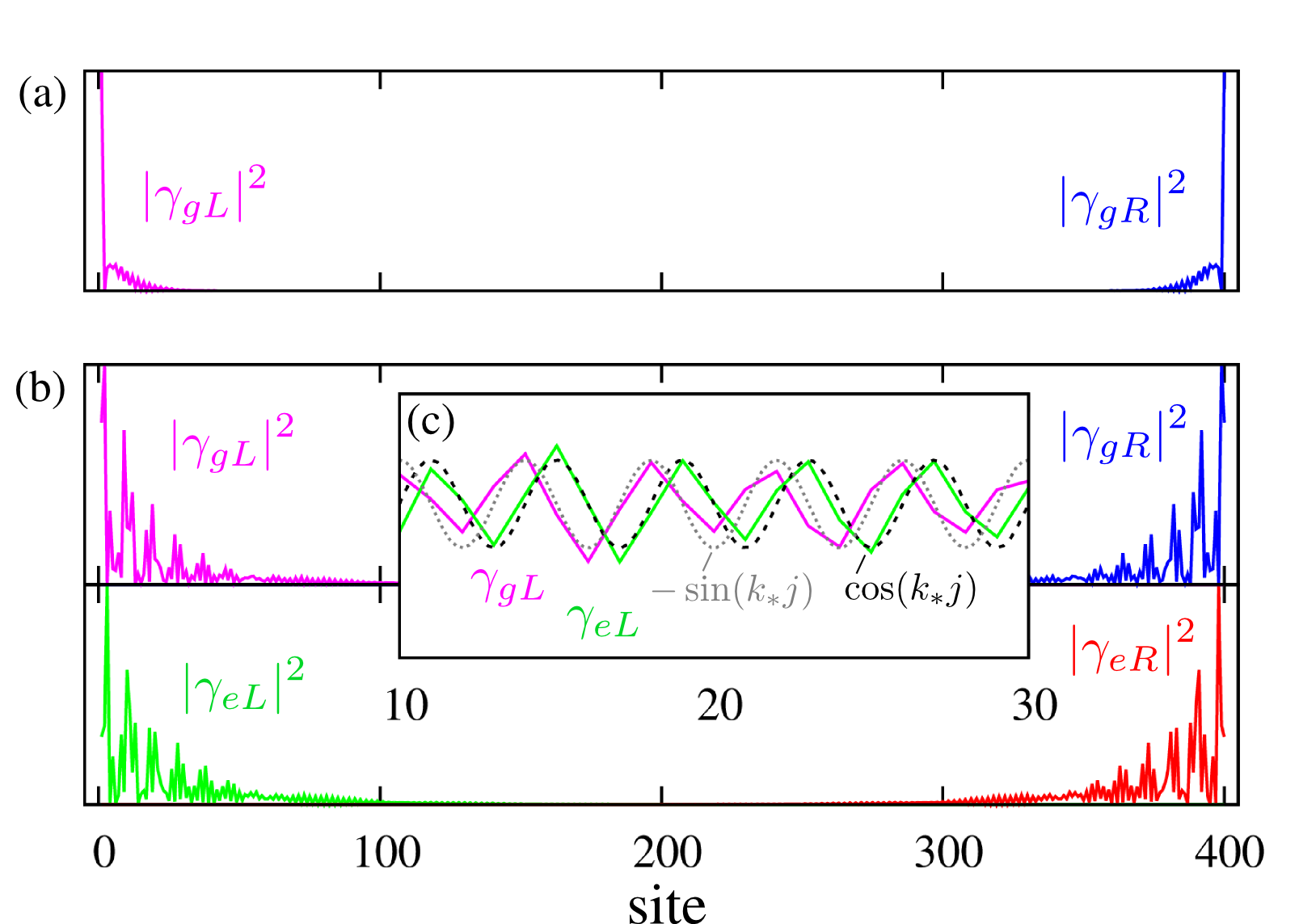}
\end{center}
\caption{
Wavefunctions corresponding to Fig.~\ref{fig::fm_finite} for the ground state (g) and the first excited state (e). The left and right Majorana bound states are labeled by
$(L)$ and $(R)$, respectively. (a) Whereas the MF wavefunction within the 1MF phase is strongly localized and oscillates with a periodicity of lattice spacing, panel (b) shows
the wavefunctions inside the 2MF phase which strongly leak into the bulk and oscillate with a wavenumber $k_*$.
}
\label{fig::wave_fcts_fm}
\end{figure}

In Fig.~\ref{fig::fm_finite} we compare the winding number calculation shown in panel (a), with the evolution of the two lowest positive eigenenergies, shown in panel (b)
that was obtained from the open chain Hamiltonian for different lengths. As follows from bulk-boundary correspondence, the number of MF bound states agrees with the value
of ${\cal N}$, although long chains are required here in order to obtain quantitative accordance with the predicted phase boundaries. As a matter of fact, this is the case for
the gap closing that occurs at the transition from trivial to ${\cal N}=2$. Here, two truly-zero energy bound states appear only for very long chains. This has to be
contrasted with the region where ${\cal N}=1$. There, the zero energy bound state become stabilized already for shorter lengths of the chain, which can be seen through the
different decays of the wavefunctions in Fig.~\ref{fig::wave_fcts_fm}(a) and (b). 

To shed more light on the above findings, we complementary demonstrate in Fig.~\ref{fig::fm_finite}(c) the gap closings of the bulk band structure, for the parameters where
the topological quantum phase transitions occur. One observes that the phase transition involving a single MF corresponds to gap closings at the inversion symmetric wavevector
$k=0$, whereas in the case involving 2MFs, the dispersion shows gap closings at two non-inversion-symmetric points $\pm k_*$.

For even better understanding, let us investigate in more detail the behavior of the topological invariant. The gap closing conditions and therefore the phase boun\-daries,
are given by setting ${\rm Det}[A_k]=0$, which requires the following two equations to be simultaneously satisfied 
\bea
v_k(\Delta+{\cal D}_k)-t_kd_k&=&0\,,\label{eq::FMcriterion1}\\t_k^2+(\Delta+{\cal D}_k)^2-{\cal B}^2-d_k^2-v_k^2&=&0\,.\label{eq::FMcriterion2}
\eea

\noi To obtain some analytical results, we will focus on a simplified situation. For instance, by considering a short superconducting coherence length, $\xi_0$, we can
restrict ourselves only up to nearest neighbor terms in the Hamiltonian of Eq.~\eqref{eq::HFMk}. Under these conditions we have $t_k=t_1\cos (ka)$, $v_k=v_1\sin(ka)$,
${\cal D}_k={\cal D}_1\cos(ka)$ and $d_k=d_1\sin(ka)$. Eq.~\eqref{eq::FMcriterion1} is satisfied for the inversion symmetric points $k=0,\pi$ and the pair of
non-inversion-symmetric points $\pm k_*$, given by $\cos(k_*a)= v_1\Delta/(d_1t_1-v_1{\cal D}_1)$. By setting these $k$-values in Eq.~\eqref{eq::FMcriterion2}, we obtain the
gap closing conditions, or equivalently the phase boundaries. For $k=0,\pi$ we obtain the condition $t_1^2+(\Delta\pm{\cal D}_1)^2={\cal B}^2$, akin to the criteria found in
nanowire models \cite{Sau,Oreg}. A similar procedure can provide the gap closing conditions for $\pm k_*$ points, which however is quite lengthy and will not be
presented here. 

It is important to comment on the form of the wavefunctions in the case of 2 MFs. In this case, the non-inversion-symmetric points $\pm k_*$, will give rise to zero-energy
wavefunctions, which however are complex and proportional to $e^{\pm ik_*aj}$ ($j$ index of chain site). Nonetheless, MF wavefunctions should be real, and this can only
achieved by making linear combinations of the wavefunctions, so that they finally obtain a dependence $\cos(k_*aj)$ and $\sin(k_*aj)$. This explains \bt{i.} the oscillating
behavior shown in Fig.~\ref{fig::wave_fcts_fm} with a period determined by $k_*$ and \bt{ii.} the fact that when one MF wavefunction shows a maximum, the other shows a
minimum. As expected, the MF wavefunction for a single MF phase does not show this type of feature.

\section{Topological AFM Shiba chain} \label{section::AFM}

The presence of Rashba SOC due to the superconduc\-ting substrate, is an ingredient capable of engineering MFs also for other magnetic phases of the chain, apart from the FM
one. As we already discussed in previous paragraphs, the same conditions which favor the FM ordering, also provide fertile ground for the establishment of AFM ordering. A
crucial requirement is the presence of strong Ising anisotropy in order to overcome the DM interaction. If this is the case, other details such as the adatom spacing, will
decide on the FM or AFM type of ordering. 

The possibility of topological AFM Shiba chains was recently discussed in Ref.~\cite{Heimes} for superconducting substrates in the absence of SOC. In that case, a new
me\-cha\-nism for engineering topological superconductivity was proposed, were SOC was induced by a supercurrent flow along the chain together with an in plane Zeeman field.
Note that AFM and FM Shiba chains can only host MFs in the presence of some kind of SOC \cite{KotetesClassi}. In stark contrast, spiral Shiba chains do not require SOC, but
exhibit MF in a self-tuned manner. Nonetheless, any realistic manipulation of MFs in spiral Shiba chains will una\-voi\-dably require the application of external Zeeman fields
\cite{JianLi} or supercurrents \cite{Ojanen}, counterbalancing the advantage of self-tunability. 

In the rest of the manuscript we will focus on the topological phases supported in AFM Shiba chains. As we will present in the next paragraphs, some of the generic results
which we reported earlier for the FM case are also relevant for the AFM order. For instance, AFM Shiba chains also support MF phases with one or two MFs per chain edge.
However, as we explain in the AFM analysis, the underlying mechanism and the topologically relevant $k$-space points, differ in each case. The reason can be traced back to the
\bt{i.} different magnetic wavevector $Q=0$ or $\pi$ and \bt{ii.} the presence of additional \textit{hidden} symmetries which appear in the AFM case.

\subsection{Symmetry classification}

The present paragraph follows closely the analysis carried out for the FM case. As previously, the relevant point group in the absence of magnetism is $C_{2v}$. The
Hamiltonian of Eq.~\eqref{eq::Hamiltonian_AFM} includes now the AFM term, given by $M_j\tau_z\sigma_z=M(-1)^j\tau_z\sigma_z$. The latter AFM Hamiltonian term, transforms under
the $C_{2v}$ elements in the following manner: $\hat{\sigma}^{\dag}_{yz}(-1)^j\tau_z\sigma_z\hat{\sigma}_{yz}=-(-1)^j\tau_z\sigma_z$,
$\hat{\sigma}^{\dag}_{xz}(-1)^j\tau_z\sigma_z\hat{\sigma}_{xz}=-(-1)^j\tau_z\sigma_z$ and $\hat{C}_2^{\dag}(-1)^j\tau_z\sigma_z\hat{C}_2=(-1)^j\tau_z\sigma_z$. Essentially, 
we recover exactly the same behavior en\-coun\-tered in the FM case, since ${\cal I}(-1)^j=(-1)^{-j}=(-1)^j$. However, in contrast to the FM ordering, the AFM chain is
invariant under the translation operation, $t_{2a}$, instead of $t_{a}$. Thus the reduced Brillouin zone (RBZ) becomes now relevant, defined by $k\in(-\pi/2a,\pi/2a]$. This
reflects the formation of a two sublattice structure. With the help of the translation operator, $\hat{t}_a$, we additionally obtain
$\hat{t}_a^{\dag}(-1)^j\tau_z\sigma_z\hat{t}_a=-(-1)^j\tau_z\sigma_z$. Finally, similarly to the FM case, the system is not invariant under ${\cal T}$, as 
$\hat{{\cal T}}^{\dag}(-1)^j\tau_z\sigma_z\hat{{\cal T}}=-(-1)^j\tau_z\sigma_z$. 

As in the FM case, the AFM chain is invariant under the hidden symmetry operator $\hat{\Theta}\equiv\hat{\sigma}_{xz}\hat{{\cal T}}=\hat{{\cal K}}$. More importantly, the
distinct property $\hat{t}^{\dag}_a(-1)^j\tau_z\sigma_z\hat{t}_a=-(-1)^j\tau_z\sigma_z$ can yield additional hidden symmetries, when $t_a$ is combined with $\sigma_{yz}$,
$\sigma_{xz}$ or ${\cal T}$. Indeed we find three additional symmetries: \bt{i.} the anti-unitary symmetry $\Theta'={\cal T}t_a$ \bt{ii.} the unitary symmety
${\cal O}=\sigma_{xz}t_a$ and \bt{iii.} the unitary symmety ${\cal O}'=\sigma_{yz}t_a$. On the other hand, unitary symmetries allow to block-diagonalize the Hamiltonian and
label it with the eigenvalues of the respective operators. Here we may use only one of the two unitary symmetry operators for block diagonalizing the Hamiltonian. Note that
the presence of two anti-unitary symmetries $\Theta$ and $\Theta'$, does not allow the classification of the Hamiltonian according to the ten existing symmetry classes
\cite{Altland,KitaevClassi,Ryu}. The latter classification methods can be \textit{only} applied on Hamiltonians with no additional unitary symmetries present. However, after
the block dia\-go\-na\-li\-za\-tion of the Hamiltonian relying on the unitary symmetry, a symmetry classification is possible \cite{KotetesClassi}. This is exactly the tactic
which we will follow in the next paragraph, by first transferring to the RBZ.

\subsection{AFM Shiba chain Hamiltonian}

By transferring to momentum space, we obtain the following Schr\"odinger equation, which provides the single-particle spectrum in the AFM case:
\bea
{\cal H}_k^0\phi_k+{\cal H}_k^{\rm m}\phi_{k+Q}=\varepsilon\phi_k\,,
\eea

\noi with $Q=\pi/a$, $k\in$ BZ, ${\cal H}_k^0$ and ${\cal H}_k^{\rm m}$ given in Eq.~\eqref{eq::HFMk}. By passing to the RBZ we obtain 
\bea
\left(\begin{array}{cc}
{\cal H}_{k-Q/2}^0&{\cal B}\tau_z\sigma_z\\
{\cal B}\tau_z\sigma_z&{\cal H}_{k+Q/2}^0\end{array}\right)\left(\begin{array}{c}\phi_{k-Q/2}\\\phi_{k+Q/2}\end{array}\right)=
\varepsilon\left(\begin{array}{c}\phi_{k-Q/2}\\\phi_{k+Q/2}\end{array}\right)\,,\quad
\eea

\noi where by additionally introducing the $\bm{\rho}$ Pauli matrices in the AFM space we end up with the Hamiltonian
\bea
\widetilde{\cal H}_k={\cal H}_{k,+}^0+{\cal H}_{k,-}^0\rho_z+{\cal B}\tau_z\rho_x\sigma_z
\eea

\noi defined in the RBZ, while we introduced
\bea
{\cal H}_{k,\pm}^0=\frac{{\cal H}_{k-Q/2}^0\pm{\cal H}_{k+Q/2}^0}{2}\,.
\eea

\noi The explicit form reads
\begin{align}
\widetilde{\cal H}_k&=t_{k,+}\tau_z+t_{k,-}\tau_z\rho_z-v_{k,+}\tau_z\sigma_y-v_{k,-}\tau_z\rho_z\sigma_y\no\\
&+(\Delta+{\cal D}_{k,+})\tau_y\sigma_y+{\cal D}_{k,-}\tau_y\rho_z\sigma_y-d_{k,+}\tau_y-d_{k,-}\tau_y\rho_z\no\\
&-{\cal B}\tau_z\rho_x\sigma_z\,,
\end{align}

\noi where the parameters appearing can be directly retrieved by the definitions of $t_k$, $v_k$, ${\cal D}_k$ and $d_k$. For completeness, we present their expression below
\bea
t_{k,+}&=&\sum_{l=1}^{\infty}t_{2l}\cos(2lka)(-1)^l\,,\\ 
t_{k,-}&=&\sum_{l=1}^{\infty}t_{2l-1}\sin[(2l-1)ka](-1)^l\,,\\
v_{k,+}&=&\sum_{l=1}^{\infty}v_{2l}\sin(2lka)(-1)^l\,,\\
v_{k,-}&=&\sum_{l=1}^{\infty}v_{2l-1}\cos[(2l-1)ka](-1)^{l+1}\,,\\
{\cal D}_{k,+}&=&\sum_{l=1}^{\infty}{\cal D}_{2l}\cos(2lka)(-1)^l\,,\\
{\cal D}_{k,-}&=&\sum_{l=1}^{\infty}{\cal D}_{2l-1}\sin[(2l-1)ka](-1)^l\,,\\
d_{k,+}&=&\sum_{l=1}^{\infty}d_{2l}\sin(2lka)(-1)^l\,,\\
d_{k,-}&=&\sum_{l=1}^{\infty}d_{2l-1}\cos[(2l-1)ka](-1)^{l+1}\,.
\eea

\noi At this point, we move on with the symmetry classification. In the particular basis, the translation operator $\hat{t}_a$ has the representation
\bea
\hat{t}_a=\left(\begin{array}{cc} e^{i(k-Q/2)a}&0\\0&e^{i(k+Q/2)a}\end{array}\right)=-i\rho_ze^{ika}\,.
\eea

\noi For simplicity, we will drop the $U(1)$ phase factor, since it is irrelevant for the present discussion. On the other hand, complex conjugation has the following
representation in this basis $\hat{\cal K}=\rho_x\hat{\cal K}'$, with $\hat{\cal K}'$ not acting on the wavevector $Q$. Under these conditions we obtain the representation for
the following operators: $\hat{\Theta}=\rho_x\hat{\cal K}'$, $\hat{\Theta}'=\rho_y\sigma_y\hat{\cal K}'$ and $\hat{\cal O}=\rho_z\sigma_y$. We directly confirm that the
Hamiltonian is invariant under the action of these operators, as discussed in the previous paragraph. However, there are additional symmetries. We find two chiral symmetries:
$\hat{\Pi}\equiv\tau_x$ and $\hat{\Pi}\equiv\tau_x\rho_z\sigma_y$, as also two charge-conjugation symmetries: $\hat{\Xi}\equiv\tau_x\rho_x\hat{\cal K}'$ and
$\hat{\Xi}'\equiv\tau_x\rho_y\sigma_y\hat{\cal K}'$. In this representation both time-reversal symmetry operators satisfy $\hat{\Theta}^2=(\hat{\Theta}')^2=I$, yielding the
symmetry class BDI$\oplus$BDI.

The particular symmetry class of the Hamiltonian can alternatively retrieved by block diagonalizing the Hamiltonian via the transformation
\bea
U=\frac{\rho_y + \rho_z}{\sqrt{2}}\frac{\rho_z\sigma_z+\sigma_y}{\sqrt{2}}e^{-i\tfrac{\pi}{4}\sigma_y}\,,
\eea

\noi which yields $U \widetilde{{\cal H}}_{k} U^\dagger=\tfrac{1}{2}\sum_\sigma (1+\sigma\sigma_z)\otimes \widetilde{\cal H}_{k,\sigma}$, with the blocks
\begin{align}
\label{eq::Hamiltonian_momentum_wire}
&\widetilde{{\cal H}}_{k,\sigma}=(t_{k,+}-\sigma v_{k,-})\tau_z+(t_{k,-}-\sigma v_{k,+})\tau_z\rho_y+{\cal B}\tau_z\rho_z\no\\
&+[\sigma(\Delta+{\cal D}_{k,+})-d_{k,-}]\tau_y\rho_y+(\sigma{\cal D}_{k,-}-d_{k,+})\tau_y\,.
\end{align}

Interestingly we find that for each subspace, $\sigma$ up and down, the Hamiltonian possesses the form of two decoupled topological FM Shiba chain models (see
Eq.~\eqref{eq::HFMk}), but with the AFM Pauli matrices playing the role of the spin Pauli matrices. Note, that with the particular choice of the spinor, the functions in
front of the matrices have a similar behavior under inversion ($k\rightarrow-k$), as in the FM Shiba case studied earlier or related nanowire models \cite{Sau,Oreg,Tewari and
Sau,Hui2014}. Therefore, we anticipate at least an equally rich phase diagram, exhibiting an interplay of topological phases with one or two MFs per edge of the chain.

\subsection{Topological invariant}

Each of the $\sigma$ subblocks reside in the BDI symmetry class and can be off-block diagonalized, similar to the procedure followed in the FM case. Therefore, we effect the
transformation $(\tau_z+\tau_x)/(\sqrt{2})$ which yields 

\begin{align}
\widetilde H_{k,\sigma}'=\begin{pmatrix}0 & A_{k,\sigma}\\A_{k,\sigma}^{\dag}&0\end{pmatrix}\,,
\end{align}

\noi with the upper off-diagonal block given by
\bea
A_{k,\sigma}=t_{k,+}-\sigma v_{k,-}+i(d_{k,+}-\sigma{\cal D}_{k,-})+{\cal B}\rho_z\no\\
-\big\{\sigma v_{k,+}-t_{k,-}-i[\sigma(\Delta+{\cal D}_{k,+})+d_{k,-}]\big\}\rho_y\,.
\eea

By introducing the determinants ${\rm Det}[A_{k,\sigma}]$, as also the related vectors $\bm{g}_{k,\sigma}=({\rm Re}~{\rm Det}[A_{k,\sigma}],
{\rm Im}~{\rm Det}[A_{k,\sigma}],0)$, we can define the quantities
\bea \label{eq::winding_2}
{\cal N}_{\sigma}=\frac{1}{2\pi}\int_{RBZ} dk \, \left(\hat{\bm{g}}_{k,\sigma}\times \frac{\partial\hat{\bm{g}}_{k,\sigma}}{\partial k}\right)_z\,,
\eea

\noi with the unit vectors $\hat{\bm{g}}_{k,\sigma}=\bm{g}_{k,\sigma}/|\bm{g}_{k,\sigma}|$. However, the quantities above \textit{do not} constitute topological inva\-riants
because the $\bm{g}_{k,\sigma}$ vectors are not compactified in the RBZ, i.e. do not have the same value for the RBZ edges $k=\pm\pi/2a$. The latter occurs because we
chose to work in the AFM space $\{k-Q/2,k+Q/2\}$, instead of the band-index space. Only Hamiltonians defined in the band index space satisfy the compactification condition.
In the present situation, the folding of $k$-space has been performed in a convenient manner, which however does not meet the above criterion. Therefore, a to\-po\-lo\-gi\-cal
invariant can be only defined by combining the two $\sigma$ sectors. Essentially we have to start from the total Hamiltonian $\widetilde{\cal H}_{k}$, block off-diagonalize
it, introduce the upper off-diagonal block $\tilde{A}_{k}$ and define a corresponding vector $\tilde{\bm{g}}_{k}=({\rm Re}~{\rm Det}[\tilde{A}_{k}],
{\rm Im}~{\rm Det}[\tilde{A}_{k}],0)$. This procedure yields the topologically invariant quantity
\bea
{\cal N}={\cal N}_{\uparrow}+{\cal N}_{\downarrow}\label{eq::winding}
\eea

\noi with ${\cal N}_{\sigma}$ being $\mathbb{R}$, instead of $\mathbb{Z}$. Note this procedure was circumvented in Ref.~\cite{Heimes} by extending the integration to the BZ.
However, the method presented in this paragragh is the most general and we conclude that \textit{only} ${\cal N}$ is capable of providing the related $\mathbb{Z}$ number of
MFs per edge which are protected by chiral symmetry. 

Nonetheless, there can be situations where additional terms in the Hamiltonian can violate chiral symmetry while at the same time preserving the unitary symmetry ${\cal O}$.
In this case, each Hamiltonian block $\widetilde{\cal H}_{k,\sigma}$ belongs to symmetry class D, which is characterized by a strong $\mathbb{Z}_2$ invariant in one momentum
space dimension. However, due to interdependence of the two blocks, only phases with 0 or 1 MFs are accessible. The phase diagram is retrieved by introducing a total
$\mathbb{Z}_2$ invariant obtained by muliplying the $\mathbb{Z}_2$ invariants of each block.

\subsection{Topological phase diagram -- Results}

\begin{figure}[t]
\begin{center}
\includegraphics[width=\columnwidth]{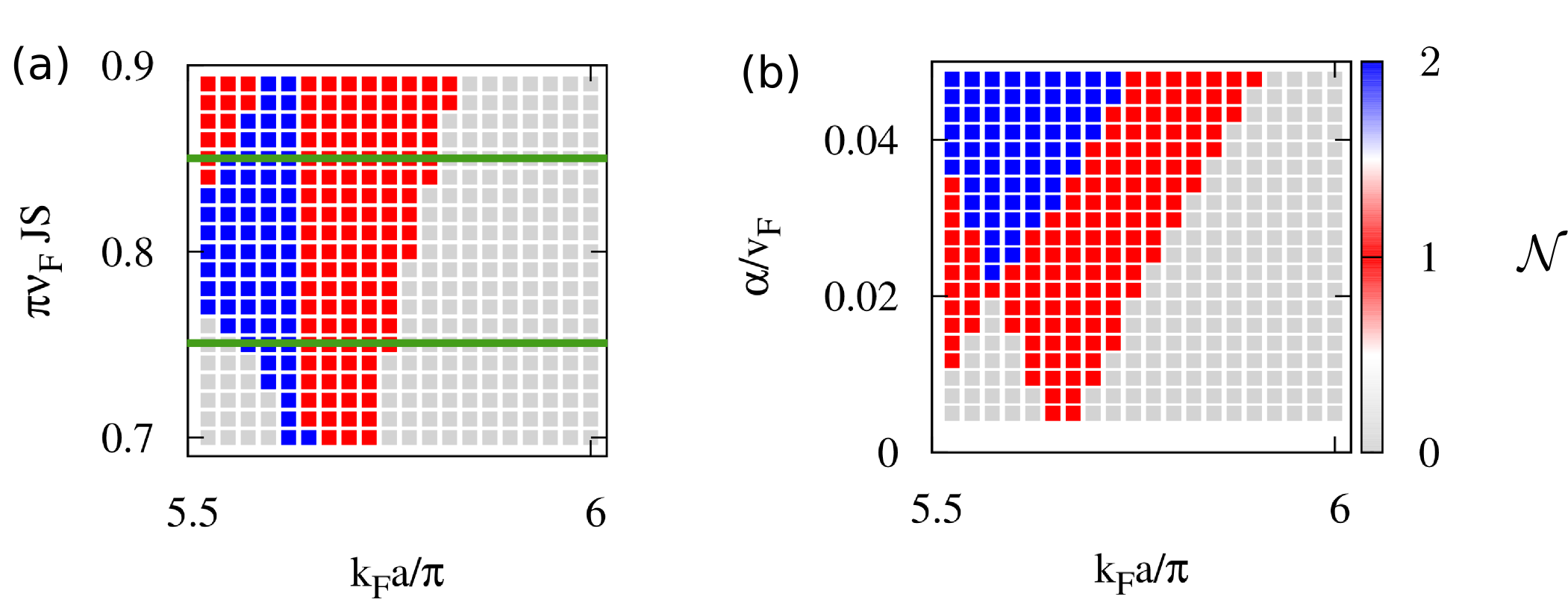}
\end{center}
\caption{The $\mathbb{Z}$ topological invariant (winding number) as defined in Eq.~\eqref{eq::winding}, for varying adatom spacing $a$ and (a) magnetic exchange energy $JS$
($\alpha=0.03\,v_F$) or (b) norma\-li\-zed SOC strength $\alpha$ ($\pi\nu_FJS=0.85$). In both cases, we find topological phases harboring 1 or 2 MFs per chain edge.
Note in (a), that tuning the magnetic exchange energy can be used to switch between 1 and 2 MF phases. When close to the phase boundary, this could achieved with a weak
perpendicular Zeeman field ($x$ axis). Observe also that a threshold SOC strength is generally required for both 1 and 2 MF phases to become accessible. This is in contrast to
the FM case and arises because the strength for the SOC $\alpha$ appears now in the topological criterion for the 1MF phase. Therefore, also tu\-ning of $\alpha$ can be
exploited for realizing topological quantum phase transitions, but via a different mechanism.}
\label{fig::fig_alpha_a}
\end{figure}

In Fig.~\ref{fig::fig_alpha_a} we present the calculated winding number ${\cal N}$ of Eq.~\eqref{eq::winding}, with varying adatom spacing $a$ and (a) magnetic exchange
energy $\pi\nu_F JS$ or (b) normalized SOC strength $\alpha$. As in the FM case, we also encounter phases with zero, one or two MFs per edge. The modification of the magnetic
exchange energy, effected for instance by applying of a Zeeman field perpendicular to the ordered spins ($x$ axis), can tune the phase diagram. Similar func\-tionality appears
with the variation of the SOC strength, where its increase can extend the window for phases with 2 MFs.

However, in contrast to the FM case, we observe that \textit{generally} a critical SOC strength is required for realizing a transition to the topological phases. The latter
feature will be explained below, by considering a nearest neighbor model for the AFM Shiba chain. In short, the apparent difference relies on the fact that for the 1MF phase
of the FM and AFM chains, different $k$ points are topologically involved. For the FM case, the inversion-symmetric points $k=0,\pi$ become relevant. In contrast, the 1MF
phase in the AFM case arise from gap closings of each $\sigma=\uparrow,\downarrow$ block Hamiltonian at the $k=0$ point of the RBZ, which coincides with the $\pi/2a$ point of
the original BZ. Therefore, the topological criteria are retrieved from different points, with the SOC strength not involved in the FM case but crucially appearing in the AFM
case.  

\begin{figure}[b]
\begin{center}
\includegraphics[width=\columnwidth]{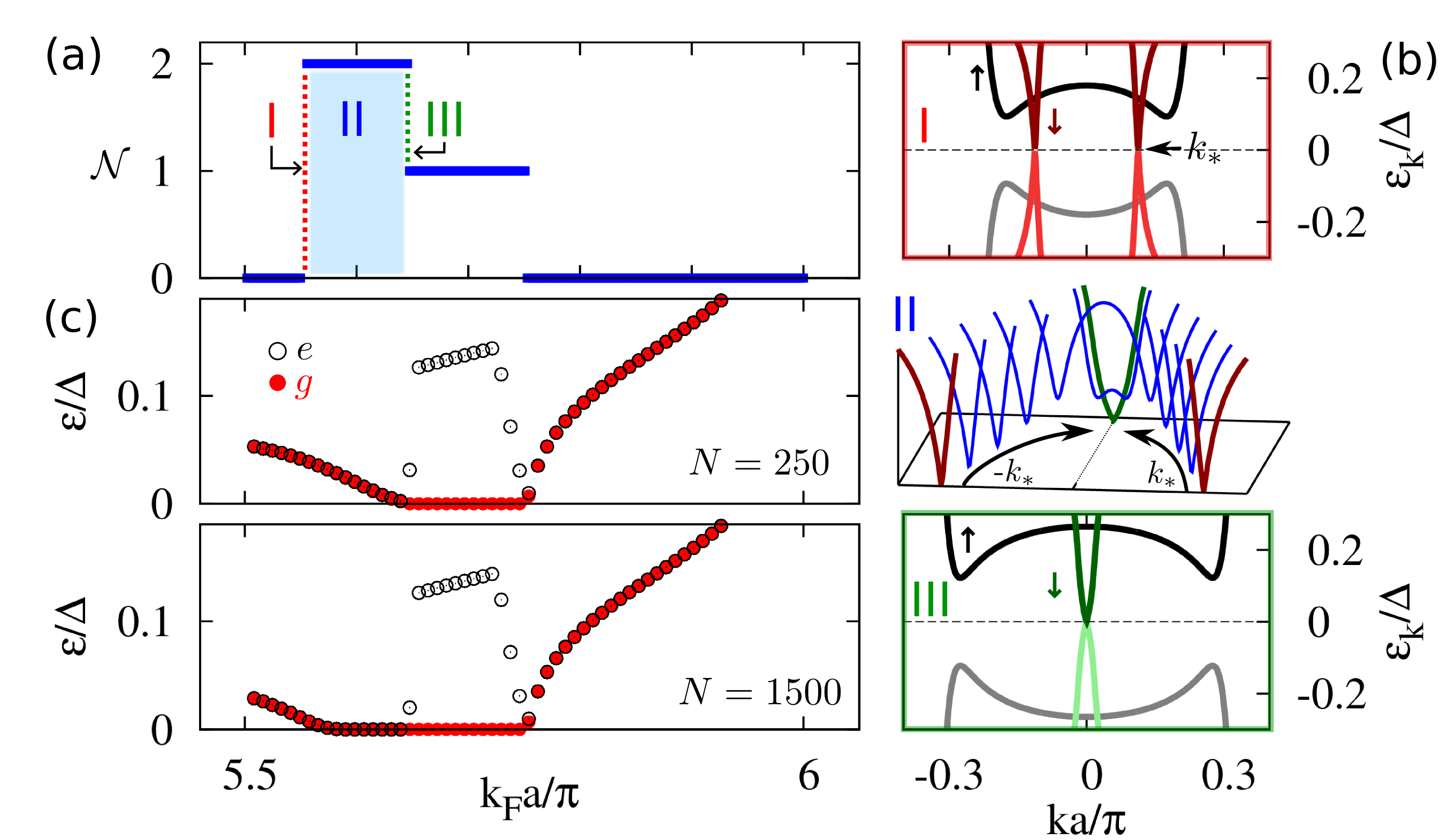}
\end{center}
\caption{(a) The invariant $\cal N$ as defined in Eq.~\eqref{eq::winding} along the green line in Fig.~\ref{fig::fig_alpha_a}(a) for $\pi\nu_FJS=0.75$. (b) We depict the
e\-ner\-ge\-ti\-cal\-ly lowest dispersions corresponding to the two blocks $\sigma=\uparrow,\,\downarrow$ of the Hamiltonian in Eq.~\eqref{eq::Hamiltonian_momentum_wire}, at
the transition points where ${\cal N}=0\rightarrow 2$ ({\color{red}I}) and ${\cal N}=2\rightarrow 1$ ({\color{darkgreen}III}). Inbetween the two critical spacings,
corresponding to ({\color{red}I}) and ({\color{darkgreen}III}), the previous gap closing points $\pm k_*$ move towards $k=0$ ({\color{blue} II}). (c) Ground state energy (red
dots) and first excited ener\-gy (black) depending on $a$ for a chain length of $N=250$ and $N=1500$ atoms.}
\label{fig::fig_inv_energies_3}
\end{figure}

We now proceed with examining in more detail the topological properties of the system for two values of the magnetic exchange energy. First we consider a cut of
Fig.~\ref{fig::fig_alpha_a} for $\alpha=0.03\,v_F$ and $\pi\nu_FJS=0.75$. In Fig.~\ref{fig::fig_inv_energies_3} we present: (a) the topological invariant ${\cal N}$ and (b)
the re\-le\-vant gap closings in RBZ associated with the changes of ${\cal N}$. We observe in Fig.~\ref{fig::fig_inv_energies_3}(b) that the transition ${\cal
N}=0\rightarrow2$ occurs due to the gap closings at the points $\pm k_*$ for $\sigma=\downarrow$. The particular phase with 2MFs is protected by chiral symmetry. Upon
increasing the adatom distance in phase II, the $\pm k_*$ points converge to $k=0$ and merge, exactly when another topological phase transition occurs ${\cal
N}=2\rightarrow1$. The latter transition and change in ${\cal N}$ is possible due to the recombination of the two $\pm k_*$ points at the inversion symmetric point $k=0$ of
the RBZ. The last transition to the trivial superconducting phase occurs via a gap closing at $k=0$ of the $\sigma=\downarrow$ subblock. Note generally that the dependence of
${\cal N}$ on the adatom spacing $a$ is quite complicated, as all the coefficients are functions of the latter. In Fig.~\ref{fig::fig_inv_energies_3}(c) we depict the
two-lowest positive eigenenergies of the AFM Hamiltonian for an open chain. Note that, the ap\-pea\-rance of a single zero eigenenergy agrees very well with the bulk
predictions for the 1MF phase. In contrast, the bulk results for the 2MF phase are retrieved for quite long chains.

\begin{figure}[t]
\begin{center}
\includegraphics[width=0.8\columnwidth]{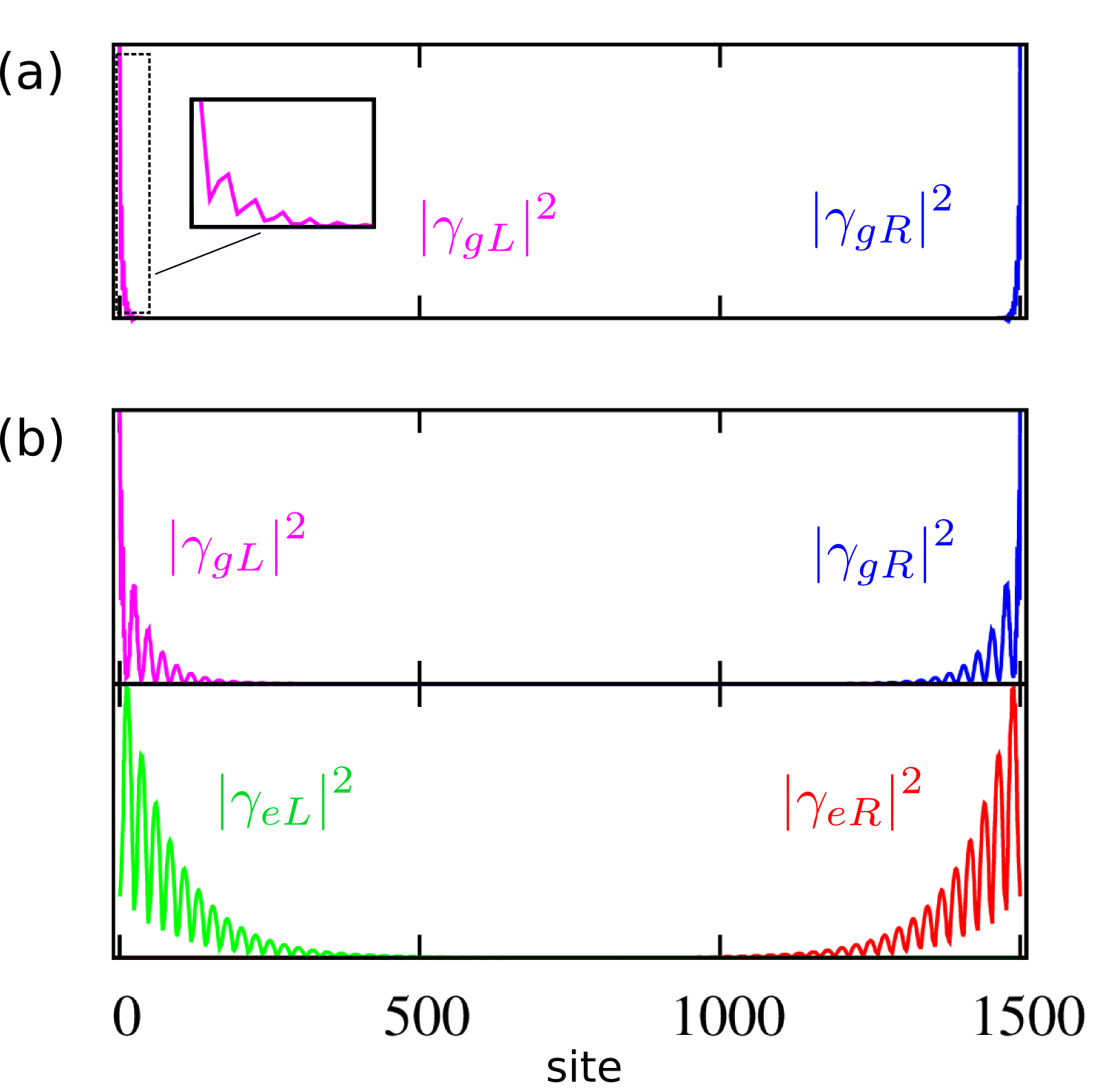}
\end{center}
\caption{Majorana wavefunctions corresponding to Fig.~\ref{fig::fig_inv_energies_3} for the ground state (g) and the first excited state (e). The left and
right Majorana bound states are labeled by $(L)$ and $(R)$ respectively. (a) The MF wavefunction in the 1MF phase shows an oscillatory dependence on the lengthscale of the 
adatom spacing. (b) The wavefunctions inside the 2MF phase are less localized and oscillate with an inverse wavelength $k_*$.} 
\label{fig::wave_fcts_afm_1}
\end{figure}

In Fig.~\ref{fig::wave_fcts_afm_1} we present the arising MF wavefunctions in the 1 MF and 2 MF cases. In the case with 2 MFs, we retrieve once again the oscillatory behavior
of the wavefunctions associated with the $\cos(k_*aj)$ and $\sin(k_*a j)$, related to chiral symmetry. However, the MF wavefunction for the 1MF phase, shows also a particular
oscillatory behaviour due to different reasons. Since the latter topological phase is arising from the $k=0$ point of the RBZ, which coincides with the $k=\pi/2a$ point of the
BZ, the wavefunctions show close to this transition point a characteristic oscillatory behavior given by the lattice constant, i.e. it assumes the form $\cos(j\pi/2)$. 
This oscillatory behavior, with a wavelength given by the adatom spacing, still persists even deep inside the 1MF phase (see Fig.~\ref{fig::wave_fcts_afm_1}(a)).

To obtain further insight, we will retrieve some ana\-lytical results by restricting to the nearest and next nea\-rest neighbor versions of the Hamiltonian in
Eq.~\eqref{eq::HFMk}. For the nearest neighbor model we have $t_k=t_1\cos (ka)$, $v_k=v_1\sin(ka)$, ${\cal D}_k={\cal D}_1\cos(ka)$ and $d_k=d_1\sin(ka)$. Each Hamiltonian
block now obtains the form
\bea
\widetilde{{\cal H}}_{k,\sigma}=-\sigma v_1\cos(ka)\tau_z-t_1\sin(ka)\tau_z\rho_y+{\cal B}\tau_z\rho_z\no\\
+[\sigma\Delta-d_1\cos(ka)]\tau_y\rho_y-\sigma{\cal D}_1\sin(ka)\tau_y\,.\label{eq::toyHam}
\eea

\noi The apparent exchange of roles between $t_1\leftrightarrow v_1$ and ${\cal D}_1\leftrightarrow d_1$ happens because the $k=0$ point of the RBZ corresponds to the
$k=\pi/2a$ of the BZ. This is exactly the reason for the distinctly different dependence on the SOC, that we obtain in the AFM topological phase dia\-grams. Therefore, gap
closings at $k=0$, connected to a 1MF phase, will occur when $v_1^2+(\sigma\Delta-d_1)^2={\cal B}^2$ depending on each $\sigma$ block. Obviously the topological phase
boundaries for the 1MF phase depends on the SOC strength, in \textit{contrast} to the FM case and nanowires proposals. 

On the other hand, the chiral symmetry protected points are given by $\cos(k_{\sigma,*}a)= \sigma t_1\Delta/(t_1d_1-v_1{\cal D}_1)$. Since $k_*\in(-\pi/2a,\pi/2a]$, we obtain
${\rm sgn}[\cos(k_{\sigma,*}a)]={\rm sgn}[\sigma]$. This implies that for each $\sigma$ block we obtain a single $k_{\sigma,*}$ satisfying the gap closing criterion. Even
more, chiral symmetry here implies that for a point $k_{\sigma,*}$, there exists another in the $-\sigma$ spin block for $k_{-\sigma,*}=-k_{\sigma,*}$. Thus the $\pm k_*$ pair
of chiral symmetry protected points found in the FM case, translates now into the $(k_{\sigma,*},k_{-\sigma,*})$ pair of points, i.e. inversion connects the two subblocks.
This also explains why we can not generally consider the quantities ${\cal N}_{\sigma}$ as independent chiral symmetry related topologically invariant quantities. 

Nonetheless, a direct comparison with the results presented in Fig.~\ref{fig::fig_inv_energies_3}, shows that a nearest neighbor model is inadequate for capturing the
physics of the exact model, since the chiral symmetry protected 2MF phase ori\-gi\-nates from $\pm k_*$ points of the same subblock. This can only occur if we take into
account the next nearest neighbor contributions. In fact, for a model with \textit{only} next nearest neighbor terms, the Hamiltonian coincides with that of
Eq.~\eqref{eq::toyHam} but with $k\rightarrow2k$ or $a\rightarrow2a$. Thus we may equivalently make use of Eq.~\eqref{eq::toyHam} but now with $k$ in the original BZ. As a
result, the equation $\cos(k_{\sigma,*}a)= \sigma t_1\Delta/(t_1d_1-v_1{\cal D}_1)$ can now provide a set of $\pm k_*$ points, for each $\sigma$ subblock, explaining our
findings. 

\begin{figure}[t]
\begin{center}
\includegraphics[width=\columnwidth]{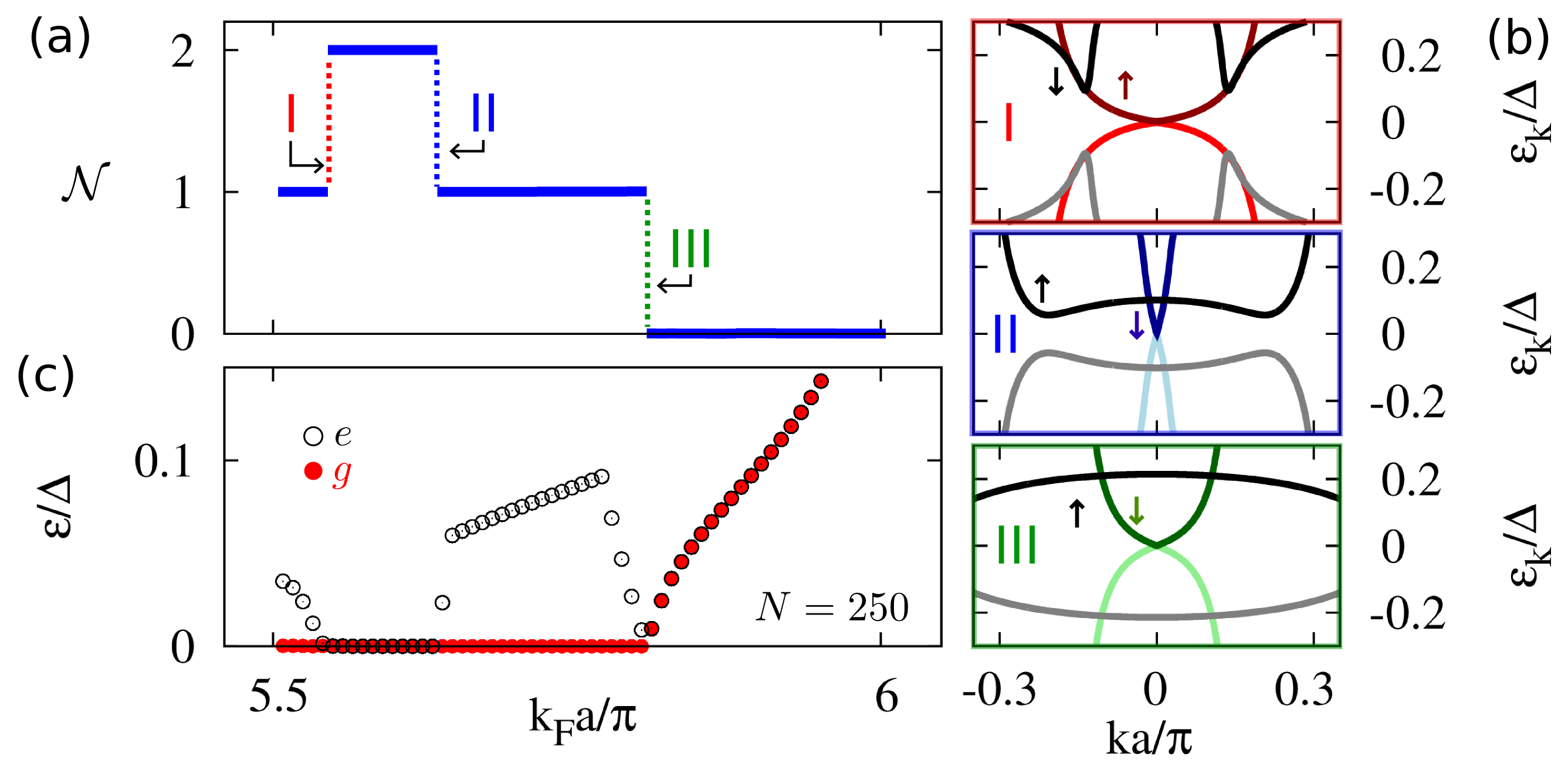}
\end{center}
\caption{
(a) The invariant $\cal N$ as defined in Eq.~\eqref{eq::winding} along the second green line in Fig.~\ref{fig::fig_alpha_a}(a) for $\pi\nu_FJS=0.85$. (b) We depict the
ener\-getically lowest dispersions corresponding to the two blocks $\sigma=\uparrow,\,\downarrow$ of the Hamiltonian in Eq.~\eqref{eq::Hamiltonian_momentum_wire}, at the
transition points where ${\cal N}=1\rightarrow 2$ ({\color{red} I}), ${\cal N}=2\rightarrow 1$ ({\color{blue} II}), and ${\cal N}=1\rightarrow 0$ ({\color{darkgreen} III}).
(c) Ground state energy (red dots) and first excited energy (black) depending on $a$ for a chain length of $N=250$ atoms. The calculations were performed for
$\alpha=0.03\,v_F$ and $\pi\nu_FJS=0.85$.}
\label{fig::fig_inv_energies}
\end{figure}

We now proceed with a cut of the phase diagram in Fig.~\ref{fig::fig_alpha_a}, for $\alpha=0.03\,v_F$ and $\pi\nu_FJS=0.85$. In Fig.~\ref{fig::fig_inv_energies} we present:
(a) the topological invariant ${\cal N}$ and (b) the relevant gap closings in RBZ associated with the changes of ${\cal N}$. For the particular value of the magnetic exchange
energy, the possibility of 2MF phases still appears, but has a different origin. This is clearly reflected in the fact that the topological invariant changes always by $1$.
This implies that only inversion-symmetric point $k=0$ can yield gap closings. This is indeed the case, as shown in Fig.~\ref{fig::fig_alpha_a}(b). We find that the different
transitions occur due to the gap closings at the $k=0$ for the two different $\sigma=\uparrow,\downarrow$ sub-blocks. Interestingly we observe that only after two successive
gap closings at $k=0$ for $\sigma=\uparrow$, the system becomes topologically trivial. This counter intui\-tive result can be naturally explained when next nearest neighbors
are present, leading to a \textit{quadratic} gap clo\-sing at $k=0$ \cite{Sun}. Essentially, the 2MF phase also in the present case, constitutes a manifestation of chiral
symmetry. 

\begin{figure}[t]
\begin{center}
\includegraphics[width=0.8\columnwidth]{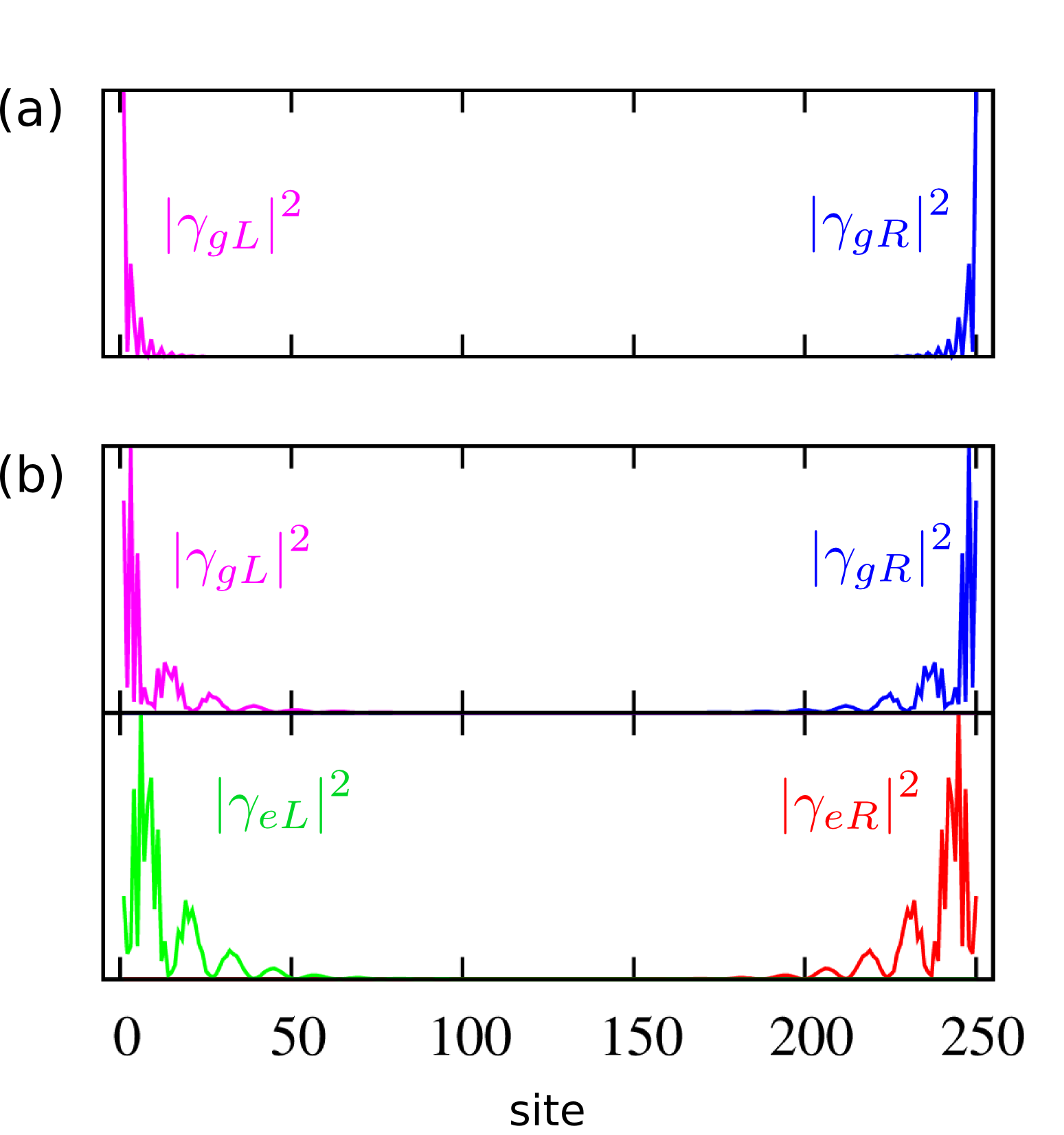}
\end{center}
\caption{Majorana wavefunctions corresponding to Fig.~\ref{fig::fig_inv_energies} for the ground state (g) and the first excited state (e). The left and right Majorana bound
states are labeled with $(L)$ and $(R)$ respectively. (a) The MF wavefunction in the 1MF phase shows an oscillatory dependence on the lengthscale of the adatom spacing. (b)
The wavefunctions inside the 2MF phase are less localized and oscillate with smaller frequency (see Fig.~\ref{fig::wave_fcts_afm_ill}).}
\label{fig::wave_fcts_3}
\end{figure}

\begin{figure}[b]
\begin{center}
\includegraphics[width=0.8\columnwidth]{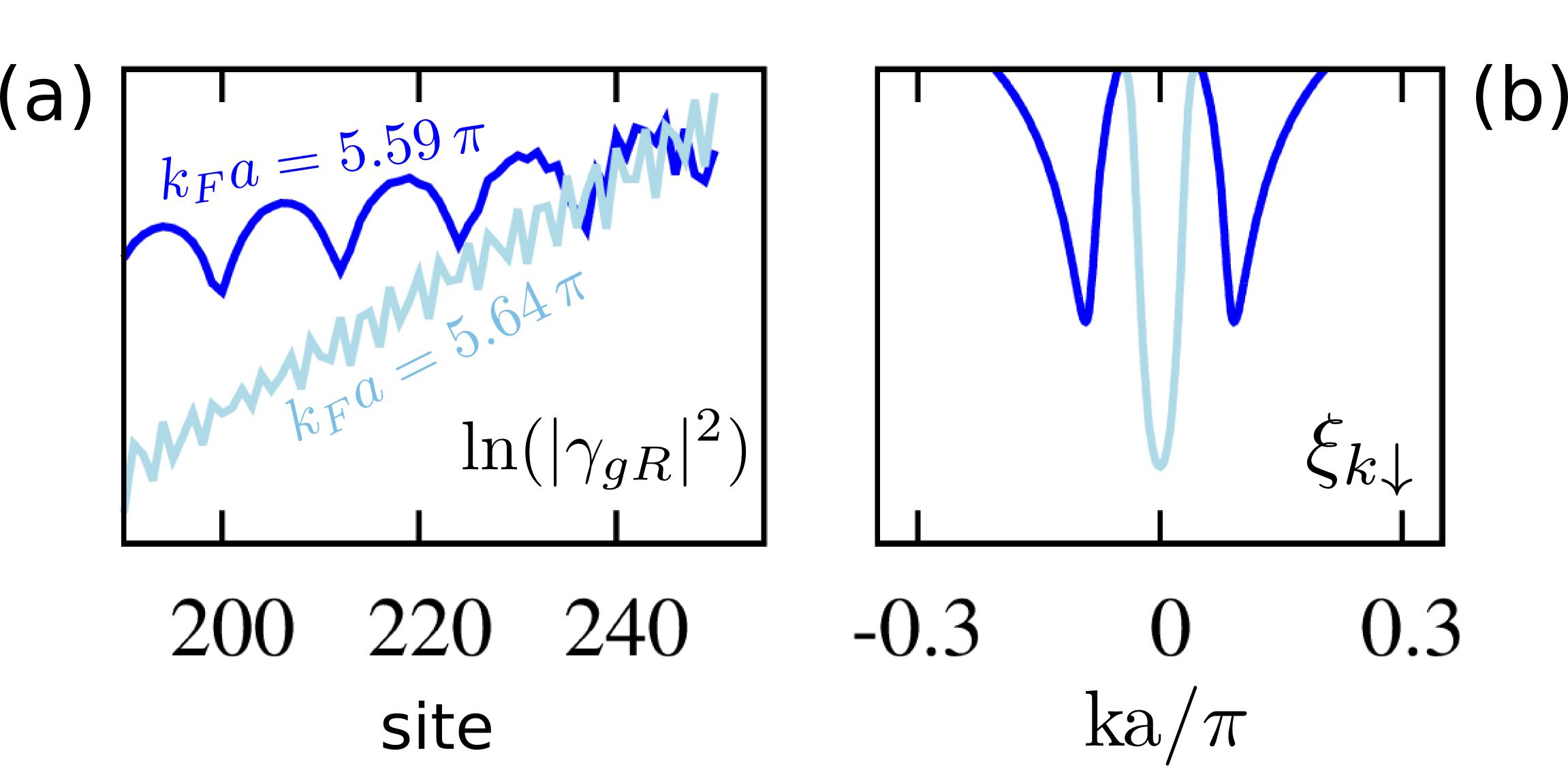}
\end{center}
\caption{(a) Logarithmic plot of the right edge MF wavefunctions: \bt{i.} deep inside the 2MF phase ($ka=5.59\,\pi$) and \bt{ii.} close to transition {\color{blue} II}
($ka=5.64\,\pi$). (b) The band minimum at $k=0$ leads to to an oscillation of the MF wavefunctions on a lengthscale of the adatom spacing (see light blue curves in (a) and
(b)). Band minima away from this point lead to the slow oscillatory trend (dark blue curves).}  
\label{fig::wave_fcts_afm_ill}
\end{figure}

In Fig.~\ref{fig::wave_fcts_3} we show representative MF wavefunctions for the ${\cal N}=1,2$ regions of Fig.~\ref{fig::fig_inv_energies}. Indeed, we find the appearance of
two MF when $k_Fa/\pi=5.59$. Each wavefunction is oscillating in magnitude and becomes exponentially suppressed in the bulk. As in previous sections, we denote the left and
right Majorana wavefunction by $\gamma_{L/R,i}$ $(i=1,2)$, respectively. Both $\gamma_L$ and $\gamma_R$ appear to be shifted spatially, with one becoming maximum at the points
where the other is minimized. Furthermore, as we show in Fig.~\ref{fig::wave_fcts_afm_ill}, the oscillatory behavior of the wavefunctions can be attributed to the band mimina
of the gapped 2MF phase. Close to transition II depicted Fig.~\ref{fig::fig_inv_energies}(a), the wavefunctions exhibit a periodicity $\cos(j\pi/2)$ which originates from the
$k=0$ point of RBZ, that coincides to the $k=\pi/2a$ point of the original BZ. In contrast, deep inside the 2MF phase and inbetween transitions {\color{red}I} and
{\color{blue}II}, the dispersion shows minima away from the $k=0$ wavevector of the RBZ, leading to oscillations with smaller frequency (see
Fig.~\ref{fig::wave_fcts_afm_ill}).

\section{Conclusions}\label{section::conclusion}

In summary, we investigated the magnetic phase diagram and the emergence of Majorana fermions in chains of magnetic adatoms deposited on a superconducting substrate with
Rashba spin-orbit coupling (SOC). By consi\-dering classical magnetic adatoms, that interact via a superexchange interaction in the additional presence of magnetic anisotropy,
we identified the parameter space for which ferromagnetic (FM), antiferromagnetic (AFM) or spiral order is stabilized. The presence of magnetic anisotropy, which arise from
the crystal field of the substrate, promotes the FM and AFM phases and renders them robust against thermal as also quantum fluctuations. 

Motivated by recent experiments which confirmed the possibility of strong magnetic anisotropy in such type of devices, we explored the occurence of topological
superconductivity for the most prominent FM and AFM configurations. Our findings reveal a rich topological phase diagram for both cases, which can support phases of 1 or 2 MFs
per edge, and can open perspectives for novel quantum computing applications. The phases with two MFs per edge are protected by chiral symmetries, which differ for each
magnetic pattern. Remarkably, the topological phase diagrams for the two cases exhibit a dif\-ferent dependence on the strength of SOC which is directly related to the
magnetic wavevector, $Q=0$ or $\pi$. In fact, depending on the value of $Q$, different points of the Shiba bandstructure become topologically relevant. As a consequence, the
MF wavefunctions demonstrate a variety of oscillatory characteristics which reflect the type of the underlying magnetic order.  

The thorough parameter exploration performed in this work, addressing \bt{i.} the competition of magnetic phases for the adatom chain and \bt{ii.} the detailed topological
phase diagram of the hybrid device, can motivate new experiments by employing alternative superconducting substrates or types of adatoms. In particular, the topological phase
diagram can be tailored via tuning the SOC strength, the adatom spacing and the magnitude of the atomic spin. Thus the emergent interplay of magnetic and therefore topological
phases in Shiba chains predicted in this work, can open the door for novel versatile and functional MF platforms.\\

\noi {\bf Note added:} {Mind that there are two regimes which describe adatom chains on top of superconductors. We discussed the Shiba {\it limit} where the
spectral weight lies entirely in the superconductor. If the adatoms are closely packed the adatom chain is in the metallic regime \cite{Li2014}. An interplay between both
regimes is most likely the situation applicable to the recent experimental results of Ref.~\cite{Yazdani_Science}. For instance, in Ref.~\cite{Peng2014} it has been shown that
a shift of spectral weight to the superconducting substrate, i.e. the Shiba limit that we considered, leads to a stronger localization of Majorana wavefunctions which is in
agreement with the recent observations \cite{Yazdani_Science}.} \\

\noi {\bf Acknowledgements:} 
We would like to thank G. Sch\"on, A. Shnirman, Y. Utsumi, G. Varelogiannis, J. Wiebe, P. M. R Brydon, A. Yazdani, S. Nadj-Perge, C. Karlewski and A. Khajetoorians for valuable 
discussions.

\appendix

\section{RKKY interaction} \label{section::appendix_RKKY} 

In this appendix we derive the effective RKKY interaction, described in Eq.~\eqref{eq::RKKY_2}, which is mediated by the electrons of a metallic surface with Rashba SOC.
We consider a chain of magnetic adatoms arranged along the $x$ direction. Accordingly, the Green's function that enters in Eq.~\eqref{eq::susceptibility}, is given by
\bea
\label{eq::calc_GF}
&&G(r\hat{\bm x},i\omega)= \frac{1}{2}\sum_{\lambda=\pm} \int \frac{d \bm k}{(2\pi)^2}\, e^{ikr \cos\varphi_{\bm{k}}}\ph \frac{1}{i\omega - \xi_{k\lambda}}\no \\
&&+ \frac{1}{2}\sum_{\lambda=\pm}\lambda\int\frac{d \bm
k}{(2\pi)^2}\,e^{ikr\cos\varphi_{\bm{k}}}\ph\frac{\sin\varphi_{\bm{k}}\sigma_x-\cos\varphi_{\bm{k}}\sigma_y}{i\omega-\xi_{k\lambda}},\no\\
&&=\frac{1}{2}\sum_{\lambda=\pm} \int_0^\infty \frac{d k\,}{2\pi} \frac{k}{i\omega - \xi_{k\lambda}}\big[J_0(kr)-i \lambda\sigma_y J_1(kr)\big]\,.\qquad
\eea

\noi Here $J_n(kr)$ are the Bessel functions which in the limit $kr\gg 1$ can be approximated by 
\begin{align}
\label{eq::bessel_approx}
J_n(kr)&\approx\sqrt{\frac{2}{\pi k|r|}} \cos\bigg(k|r|-\frac{n\pi}{2} - \frac{\pi}{4}\bigg)[{\sgn}(r)]^n\,.
\end{align}

\noi The remaining momentum integral in Eq.~\eqref{eq::calc_GF} can be derived by the substitutions $k\rightarrow k_\lambda + \xi/v_F$ and $\int_0^{\infty}
\tfrac{dk\,k}{2\pi}\rightarrow \nu_F \int_{-\infty}^{\infty} d\xi$, where $\nu_F$ is the density of states at the Fermi level. Within this approximation the remaining
integrals in \eqref{eq::calc_GF} can be evaluated, and are given by the quantities
\begin{align}
&I_n(r,i\omega)=\no\\
&\sum_{\lambda=\pm} \lambda^n\, \nu_F \int_{-\infty}^\infty d\xi\, \frac{\cos\big[(k_\lambda + \xi/v_F) |r|-\frac{n\pi}{2} - \frac{\pi}{4}\big]}{i\omega - \xi}\,,
\end{align}

\noi with $m=0,\,1$. This can be done by means of a contour integral providing
\begin{align}
\frac{I_n(r,i\omega)}{i\pi\nu_F}=-\sgn(\omega)\sum_{\lambda=\pm} \lambda^n e^{i\sgn(\omega)(k_\lambda |r|-\frac{n\pi}{2} - \frac{\pi}{4})} e^{-\tfrac{|\omega r|}{v_F}}\,.
\end{align}

\noi It follows that the electronic Green's function is ap\-proxi\-ma\-tely given by
\begin{align}
G(r\hat{\bm x},i\omega)\approx \sqrt{\frac{1}{2\pi k_F|r|}}[I_0(r,i\omega)- i \sigma_y I_1(r,i\omega)\sgn(r)]\,.
\end{align} 

\noi Note that we replaced $k$ by $k_F$ everywhere except for the arguments of the trigonometric functions. This approximation is valid as long as $\delta k \ll k_F$, which 
we assume to be the case throughout this work. In order to evaluate the susceptibility of Eq.~\eqref{eq::susceptibility}, we make use of the relation
\begin{align}
&{{\rm Tr}_{\sigma}\bigg\{\sigma_\alpha [I_0+i\sigma_yI_1\sgn(r)] \sigma_\beta [I_0-i\sigma_yI_1\sgn(r)]\bigg\}/2} \label{eq::sus}\no\\
 &= (I_0^2-I_1^2) \delta_{\alpha\beta} + 2I_1^2\delta_{\alpha,y}\delta_{\beta,y}+2\varepsilon_{\alpha\beta y}I_0I_1\sgn(r)\,.
\end{align}

\noi Furthermore in the limit $T\rightarrow 0$, the Matsubara sums in Eq.~\eqref{eq::susceptibility} can be converted into integrals, i.e. $T\sum_\omega\rightarrow
\int_{-\infty}^{+\infty} \tfrac{d\omega}{2\pi}$, yielding
\begin{align} \label{eq::integral_II}
&\int_{-\infty}^\infty { \frac{d\omega}{\pi}} I_m(r,i\omega)I_n(r,i\omega)\\
&={-\sum_{\lambda,\lambda'=\pm}} \lambda^m(\lambda')^n \sin\bigg[(k_\lambda+k_{\lambda'}) |r|-\frac{n+m}{2}\pi\bigg]\frac{\pi v_F\nu_F^2}{|r|}\,,\nonumber
\end{align}

\noi where $\nu_F=m/2\pi{}$ is the density of states for each spin-band. We use Eq.~\eqref{eq::integral_II} together with Eq.~\eqref{eq::sus} in order to evaluate the
susceptibility in Eq.~\eqref{eq::susceptibility}, which yields after some algebra the well known RKKY interaction Eq.~\eqref{eq::RKKY_2} for a two-dimensional metal with
Rashba SOC.

\end{document}